\documentclass[12pt]{article}
\pdfoutput=1
\usepackage[dvipdfmx]{graphicx}
\usepackage{color,amssymb,amsmath}
\usepackage[a4paper, total={16.5cm,22cm}]{geometry}
\usepackage{cite}
\usepackage[colorlinks=true,linkcolor=cadmiumorange,
  citecolor=electricviolet,urlcolor=electricviolet,linktocpage=false]{hyperref}

\definecolor{electricviolet}{rgb}{0.56, 0.0, 1.0}
\definecolor{cadmiumorange}{rgb}{0.93, 0.53, 0.18}

\begin{document}

\newcommand{\hf}{\mathchar`-}

\begin{titlepage}

\begin{center}
\hfill UT-18-26
\\
\hfill IPMU18-0188

\vskip 2cm

{\Large \bf 
Moduli Oscillation Induced by Reheating
}

\vskip 1.5cm

{\large Daisuke Hagihara$^{(a)}$, Koichi Hamaguchi$^{(a,b)}$ and Kazunori Nakayama$^{(a,b)}$}

\vskip 0.5cm

$^{(a)}${\em Department of Physics, 
The University of Tokyo, Tokyo 113-0033, Japan}

\vskip 0.2cm

$^{(b)}${\em Kavli IPMU (WPI), The University of Tokyo, Kashiwa 277--8583, Japan
}

\end{center}

\vskip 1.5cm

\begin{abstract}

We estimate the abundance of the coherent oscillation of moduli, which linearly couple to matter fields through higher dimensional operators.
During the (p)reheating after inflation, matter particles are efficiently produced and it can affect the moduli potential in a non-adiabatic way, which results in the coherent oscillation of the moduli.
In particular, such effects are most important at the very beginning of the (p)reheating. It is found that this production mechanism is so efficient that a successful evolution of the universe can be threatened even if the moduli mass is larger than the Hubble scale.

\end{abstract}

\end{titlepage}

\renewcommand{\thepage}{\arabic{page}}
\setcounter{page}{1}
\renewcommand{\thefootnote}{$\natural$\arabic{footnote}}
\setcounter{footnote}{0}

\newpage


\tableofcontents

\section{Introduction}
\label{sec:introduction}

Extensions of the Standard Model often possess scalar fields with relatively small masses and weak couplings to ordinary matter. 
One example is moduli, which often appear in string theoretic framework after compactification of the extra dimensions.
They are often abundantly produced in the form of coherent oscillation in the early universe.
Since moduli may have lifetime longer than $1$\,sec, their decay can threaten the success of Big-Bang nucleosynthesis as well as subsequent evolution of the universe: it is the so-called cosmological moduli problem~\cite{Coughlan:1983ci,Goncharov:1984qm,Ellis:1986zt,Banks:1993en,deCarlos:1993wie}. Even if the lifetime is much shorter than 1\,sec, the huge abundance of moduli is dangerous since the preexisting baryon asymmetry is diluted by the entropy production from the decay of moduli, and dark matter may be overproduced by the moduli decay depending on the underlying model.
Other examples of such a scalar field include saxion~\cite{Kim:1983ia,Rajagopal:1990yx}, which is a scalar partner of the axion in supersymmetric models, flavon in the Froggatt-Nielsen mechanism~\cite{Froggatt:1978nt}, Polonyi field in supergravity theories~\cite{Nilles:1983ge}, and so on. In this paper we collectively call such scalar fields as ``moduli'' and they are denoted by $\sigma$.

The abundance of coherent oscillation of the moduli may be estimated as follows. Let us first neglect the interaction of the modulus $\sigma$.
If $m_\sigma\lesssim H_{\rm inf}$, where $m_\sigma$ is the mass of the modulus and $H_{\rm inf}$ denotes the Hubble scale during inflation, the modulus is not expected to fall into the potential minimum but it in general obtains a large field value $\sigma_i$ during inflation. It begins to oscillate around the potential minimum $\sigma=0$ around when the Hubble parameter $H$ becomes close to the modulus mass $m_\sigma$.\footnote{
	Without loss of generality, we choose the zero-temperature potential minimum of the modulus as $\sigma=0$.
}
The modulus energy density $\rho_\sigma$ divided by the entropy density $s$ is estimated as
\begin{align}
	\frac{\rho_\sigma}{s} = \begin{cases}
	\displaystyle \frac{1}{8} T_R \left(\frac{\sigma_i}{M_P}\right)^2 & {\rm for}~~t_{\rm os} < t_R\\
	\displaystyle \frac{1}{8} T_{\rm os} \left(\frac{\sigma_i}{M_P}\right)^2 & {\rm for}~~t_{\rm os} > t_R
	\end{cases}
	\qquad(m_\sigma\lesssim H_\text{inf}),
	\label{abundance_usual}
\end{align}
where $M_P$ is the reduced Planck scale, $t_R$ ($T_R$) is the time (temperature) at the end of reheating and $t_\text{os}$ ($T_\text{os}$) is the time (temperature) at the beginning of the modulus oscillation.
Here we have assumed that the equation of state of the universe behaves as non-relativistic matter before the completion of the reheating. The natural value of $\sigma_i$ crucially depends on the setup. It is known that in the de-Sitter spacetime the typical field variance of the superhorizon mode becomes $\left<\sigma^2\right> \simeq 3H_{\rm inf}^4/(8\pi^2 m_\sigma^2)$ for $m_\sigma\lesssim H_{\rm inf}$ if the inflation lasts long enough~\cite{Linde:2005ht}.
However, it should not exceed the cutoff scale of the theory, which we denote by $M$. For string theoretic moduli we may have $M \sim M_P$, while for saxions or flavons $M$ may be close to the breaking scale of the Peccei-Quinn or flavor symmetry. In any case, the modulus abundance may be huge enough to threaten the cosmological history. Note that the presence of Hubble mass term does not help the situation: the Lagrangian is in general given by
\begin{align}
	\mathcal L = -\frac{1}{2}(\partial \sigma)^2 - \frac{1}{2}m_\sigma^2\sigma^2 - \frac{C}{2} H^2 (\sigma- \sigma_H)^2,
\end{align}
with some numerical coefficient $C$ and arbitrary $\sigma_H$. In such a case the amplitude becomes $\sigma_i \simeq \sigma_H$, which again results in huge modulus abundance.\footnote{
	Even in the case of $C\gg 1$~\cite{Linde:1996cx}, the modulus amplitude is not suppressed enough~\cite{Nakayama:2011wqa}.
}
The exceptional case is $\sigma_H=0$ which happens if the potential minimum $\sigma=0$ is an enhanced symmetry point, but actually it is not the case for all examples we mentioned above.

Now we want to consider the case of $m_\sigma \gtrsim H_{\rm inf}$. In such a case, the modulus falls into the minimum $\sigma=0$ during inflation and we might naively expect that no coherent oscillation is induced thereafter and hence there is no cosmological moduli problem. It is true in the limit of exactly free scalar field, but turning on the small interaction terms dramatically changes the situation. Throughout this paper, we consider the linear moduli coupling like
\begin{align}
	\mathcal L = \frac{\sigma}{M}\mathcal O,
	\label{linearterm}
\end{align}
where the dimension-four operator $\mathcal O$ consists of the other fields than $\sigma$ which may include the Standard Model fields. Moduli in general have such couplings and saxions or flavons also have such interactions after expanding the Lagrangian around their vacuum expectation values (VEVs).
During the reheating process, $\left<\mathcal O\right>$ is time dependent due to the particle production from the inflaton, and it may change the modulus potential in a non-adiabatic way. To the best of our knowledge, the modulus abundance induced in this way has not been estimated seriously.\footnote{
	The effect of linear term proportional to $T^4$ on the modulus oscillation, where $T$ denotes the temperature, was discussed in the context of moduli~\cite{Nakayama:2008ks} and flavon~\cite{Lillard:2018zts} for the case of $m_\sigma < H_{\rm inf}$. See also an earlier work~\cite{Buchmuller:2004xr}.
}
The main purpose of this paper is to study such effects qualitatively and quantitatively.

This paper is organized as follows. First, we explain the basic mechanism of moduli oscillation due to a rapid minimum shift in Sec.~\ref{sec:modulus_oscillation}. 
In Sec.~\ref{sec:example}, we consider a simple toy model as an illustrative example. The modulus abundance induced during preheating is estimated both analytically and numerically.
We summarize our study in Sec.~\ref{sec:summary}.

\section{Moduli oscillation due to linear term}
\label{sec:modulus_oscillation}

\subsection{Basic mechanism}
\label{sec:mechanism}

Let us consider a modulus, which has an interaction term in the form of Eq.~(\ref{linearterm}). In general, $\mathcal O$ can have a finite expectation value that results in the effective modulus potential
\begin{align}
	V = -\frac{\sigma}{M}\langle\mathcal O\rangle,
	\label{linear}
\end{align}
where the modulus $\sigma$ is assumed to be a real scalar field. 
Throughout this paper, we consider the case that the modulus mass $m_\sigma$ is (much) larger than the Hubble parameter $H$ during inflation:
\begin{align}
	m_\sigma \gg H_{\rm inf}.
\end{align}
The linear term shifts the position of the modulus potential minimum from the origin. The potential minimum, which we denote by $\sigma_\text{min}$, is given by\footnote{
	Note that we only consider the case of $|\sigma_{\rm min}| < M$ so that the dynamics is described within the effective theory. For moduli which only have Planck-suppressed interactions $M\sim M_P$, this condition is always satisfied. If the cutoff scale is much lower it is possible that the linear expansion around $\sigma=0$ may not be validated and we should consider the dynamics within UV-complete theory. 
}
\begin{align}
	\sigma_\text{min}=\frac{\langle\mathcal O\rangle}{Mm_\sigma^2}.
	\label{minimum}
\end{align}
The modulus cannot follow the minimum if $\sigma_\text{min}$ (or $\langle\mathcal O\rangle$) changes within a time scale shorter than $m_\sigma^{-1}$:
\begin{align}
	m_\sigma\ll\left|\frac{1}{\langle\mathcal O\rangle}\frac{d\langle\mathcal O\rangle}{dt}\right|
	\label{condition}.
\end{align}
Then the modulus starts to oscillate around the new potential minimum at some instance.
The initial oscillation amplitude is roughly given by
\begin{align}
	\hat{\sigma}(t_\text{os})\simeq\sigma_\text{min}(t_\text{os}),
	\label{sigma_relation}
\end{align}
where $\hat{\sigma}$ denotes the amplitude of the modulus oscillation and $t_\text{os}$ denotes the time at the beginning of the oscillation.
The initial energy density of the modulus oscillation is then calculated as
\begin{align}
	\rho_\sigma(t_\text{os})\simeq \frac{1}{2}m_\sigma^2 \hat{\sigma}^2(t_\text{os})
	\simeq \frac{1}{2}m_\sigma^2 \sigma_\text{min}^2(t_\text{os})
	\simeq \frac{|\Delta\langle\mathcal O\rangle|^2}{2M^2m_\sigma^2},
	\label{energy}
\end{align}
where $\Delta\langle\mathcal O\rangle$ is the variation of $\langle\mathcal O\rangle$ until $t=t_\text{os}$.

For the oscillation mechanism of our interest, both the rate and the amount of change of $\langle\mathcal O\rangle$ are important.
Even for large $|\Delta\langle\mathcal O\rangle|$, the modulus oscillation is not induced in the first place if Eq.~\eqref{condition} is not satisfied. On the other hand, even if Eq.~\eqref{condition} is fulfilled, small $|\Delta\langle\mathcal O\rangle|$ does not lead to a sizable modulus oscillation. One of the situations where Eq.~\eqref{condition} is satisfied and $|\Delta\langle\mathcal O\rangle|$ can be large at the same time is the (p)reheating era after inflation where efficient particle production happens.
In the following, we assume that there is no rapid change of $\langle\mathcal O\rangle$ after the completion of reheating.

Now let us evaluate the modulus abundance for a given $|\Delta\langle\mathcal O\rangle|$, assuming that Eq.~\eqref{condition} is satisfied. If the modulus starts to oscillate before the completion of reheating, its energy density decreases from $t_\text{os}$ to $t_R$ by a factor of $(H(t_R)/H(t_\text{os}))^2$, where we have assumed that 
the universe behaves as the matter-dominated one until the end of reheating, as is usually the case for the inflaton oscillation domination.
The modulus abundance for $t\geq t_R$ is then calculated as
\begin{align}
	\frac{\rho_\sigma}{s}
	\simeq \frac{|\Delta\langle\mathcal O\rangle|^2}{2M^2m_\sigma^2}\left(\frac{H(t_R)}{H(t_\text{os})}\right)^2\frac{45}{2\pi^2g_*T_R^3}
	\simeq \frac{T_R |\Delta\langle\mathcal O\rangle|^2}{8  H_{\rm inf}^2M_P^2 m_\sigma^2 M^2},
	\label{abundance}
\end{align}
where we have used $H(t_\text{os})\simeq H_\text{inf}$ and $3H^2(t_R)M_P^2=(\pi^2 g_*/30)T_R^4$ with $g_*$ being the effective number of degrees of freedom in thermal bath.
Note that the modulus oscillation begins within one Hubble time under our assumption $m_\sigma\gg H_\text{inf}$.
On the other hand, if the modulus starts to oscillate after the completion of reheating, the reheating is necessarily instantaneous (i.e., the reheating is completed within one Hubble time after inflation). Then, the resulting modulus abundance coincides with the last expression in Eq.~\eqref{abundance} by interpreting $3H_\text{inf}^2M_P^2=(\pi^2 g_*/30)T_R^4$. Therefore, in both cases we can use Eq.~\eqref{abundance}.
It can also be rewritten as
\begin{align}
	\frac{\rho_\sigma}{s} \simeq \frac{9T_R}{8} \left(\frac{H_{\rm inf}}{m_\sigma}\right)^2
	\left(\frac{M_P}{M}\right)^2\left(\frac{|\Delta\langle\mathcal O\rangle|}{3H_{\rm inf}^2M_P^2}\right)^2.   \label{abundance_rewritten}
\end{align}
The last factor in (\ref{abundance_rewritten}) roughly represents how efficiently the inflaton energy density is transferred to $\mathcal O$, which in principle can be as large as order unity. One can see that it is highly nontrivial whether this contribution to the modulus abundance is cosmologically safe or not.
Eq.~\eqref{abundance}, or equivalently Eq.~\eqref{abundance_rewritten}, is the formula for the moduli abundance for $m_\sigma \gg H_\text{inf}$, which is the counterpart of Eq.~\eqref{abundance_usual} for $m_\sigma \lesssim H_\text{inf}$.

Then, what we have to estimate is $|\Delta\langle\mathcal O\rangle|$. It is in general difficult to estimate it since the particle production may be highly non-equilibrium processes. In the next subsection, as a warm-up, we will consider a rather idealized situation where produced particles are instantaneously thermalized. It will be useful to capture the basic concepts of our study and how to estimate the modulus abundance.
In Sec.~\ref{sec:example}, we reconsider it with a more realistic setup.

\subsection{Example: the case of sudden transition and instant thermalization}
\label{sec:perturbative}

Let us consider the modulus Lagrangian of the form
\begin{align}
	\mathcal L = -\frac{1}{2}(\partial \sigma)^2 - \frac{1}{2}m_\sigma^2\sigma^2 + \frac{\sigma}{M}T^4, 
	\label{modulus_lagrangian}
\end{align}
where $T$ denotes the temperature.\footnote{
The effect of the last term on the moduli destabilization was discussed in Refs.~\cite{Buchmuller:2004xr,Anguelova:2009ht}.
}
This corresponds to the case of $\langle\mathcal O\rangle=T^4$ in Eq.~\eqref{linearterm}.
One of the operators to realize the situation is $\mathcal O=\chi\Box\chi$ where $\chi$ denotes a light scalar field in thermal bath and $\Box=\partial_\mu \partial^\mu$ is the d'Alembertian. 
By using a coupling strength $g_\text{th}$ parametrizing the interaction of $\chi$, we can evaluate $\langle\chi\Box\chi\rangle$ with a technique of thermal field theory as~\cite{Kawasaki:2011zi}
\begin{align}
	\langle\chi\Box\chi\rangle=\frac{m_\text{th}^2T^2}{2\pi^2}J\left(\frac{m_\text{th}}{T}\right)\simeq \kappa_\text{th}\frac{g_\text{th}^2 T^4}{12}.
\end{align}
Here $m_\text{th}^2=\kappa_\text{th} g_\text{th}^2T^2$ is the thermal mass with $\kappa_\text{th}\lesssim \mathcal O(1)$ being a model-dependent parameter, and $J(z)$ is defined by
\begin{align}
	J(z)\equiv\int_z^\infty dx\frac{\sqrt{x^2-z^2}}{e^x-1}.
\end{align}
We have used an approximation $J(m_\text{th}/T)=J(\sqrt{\kappa_\text{th}}g_\text{th})\simeq J(0)=\pi^2/6$. Therefore, by taking thermal average and correctly rescaling the cutoff scale $M$, we can reproduce the linear term in Eq.~\eqref{modulus_lagrangian}.
We obtain similar results also for the operator $\mathcal O$ consisting of fermions or gauge bosons~\cite{Dine:2000ds,Buchmuller:2003is,Buchmuller:2004xr,Kawasaki:2012qm,Kawasaki:2012rs}.  Now the effective modulus potential is temperature dependent. At first sight, this does not seem to affect the modulus abundance, since the temperature dependence is adiabatic, i.e. $|\dot T/T| \sim H \ll m_\sigma$. However, it is not always true at the very beginning of the reheating, as shown below.

Under the assumption that the inflaton decay products are instantaneously thermalized, the dynamics during reheating is described by the following equations:
\begin{align}
	&\ddot{\sigma}+3H\dot{\sigma}+m_\sigma^2\sigma-\frac{T^4}{M}=0\label{re_modulus}\\
	&\dot{\rho}_\phi+3H\rho_\phi=-\Gamma_\phi\rho_\phi\label{re_inflaton}\\
	&\dot{\rho}_\chi+4H\rho_\chi=\Gamma_\phi\rho_\phi\label{re_radiation}\\
	&3M_P^2H^2=\rho_\phi+\rho_\chi\label{re_friedmann}
\end{align}
Here $\Gamma_\phi$ denotes the perturbative decay rate of the inflaton, $\rho_\phi$ ($\rho_\chi$) is the energy density of the inflaton (the radiation) and the dot denotes the derivative with respect to the cosmic time. We have neglected the contribution of the modulus in Eq.~\eqref{re_friedmann}. Since we are assuming instant thermalization, $\rho_\chi$ is related to $T$ as $\rho_\chi=(\pi^2g_*/30)T^4$. The initial condition is chosen as $\rho_\phi(t_0)=3H_\text{inf}^2M_P^2$, $\rho_\chi(t_0)=0$ and $\sigma(t_0)=\dot{\sigma}(t_0)=0$ with $t_0$ being the time at the beginning of reheating.
At around the beginning of the evolution, noting $\rho_\chi(t)\simeq\Gamma_\phi \rho_\phi(t_0)\times (t-t_0)$, we obtain
\begin{align}
	\frac{1}{\langle\mathcal O\rangle}\frac{d\langle\mathcal O\rangle}{dt}=\frac{\dot{\rho}_\chi}{\rho_\chi}
	\simeq\frac{1}{t-t_0}.  \label{dotO/O}
\end{align}
Since this increases as $t$ comes close to $t_0$, the condition \eqref{condition} is satisfied at least for $0\leq t-t_0\ll t_\text{os}-t_0\sim m_\sigma^{-1}$. Therefore some amount of modulus oscillation is necessarily induced in the present setup.
In the following, we estimate the modulus abundance by using Eq.~\eqref{abundance} with $|\Delta\langle\mathcal O\rangle|=T_\text{os}^4$.
We classify the situation into three cases.
In all cases the modulus oscillation happens within one Hubble time after inflation.
\begin{figure}[t]
	\begin{minipage}{0.32\hsize}
		\centering
		\includegraphics[width=\linewidth]{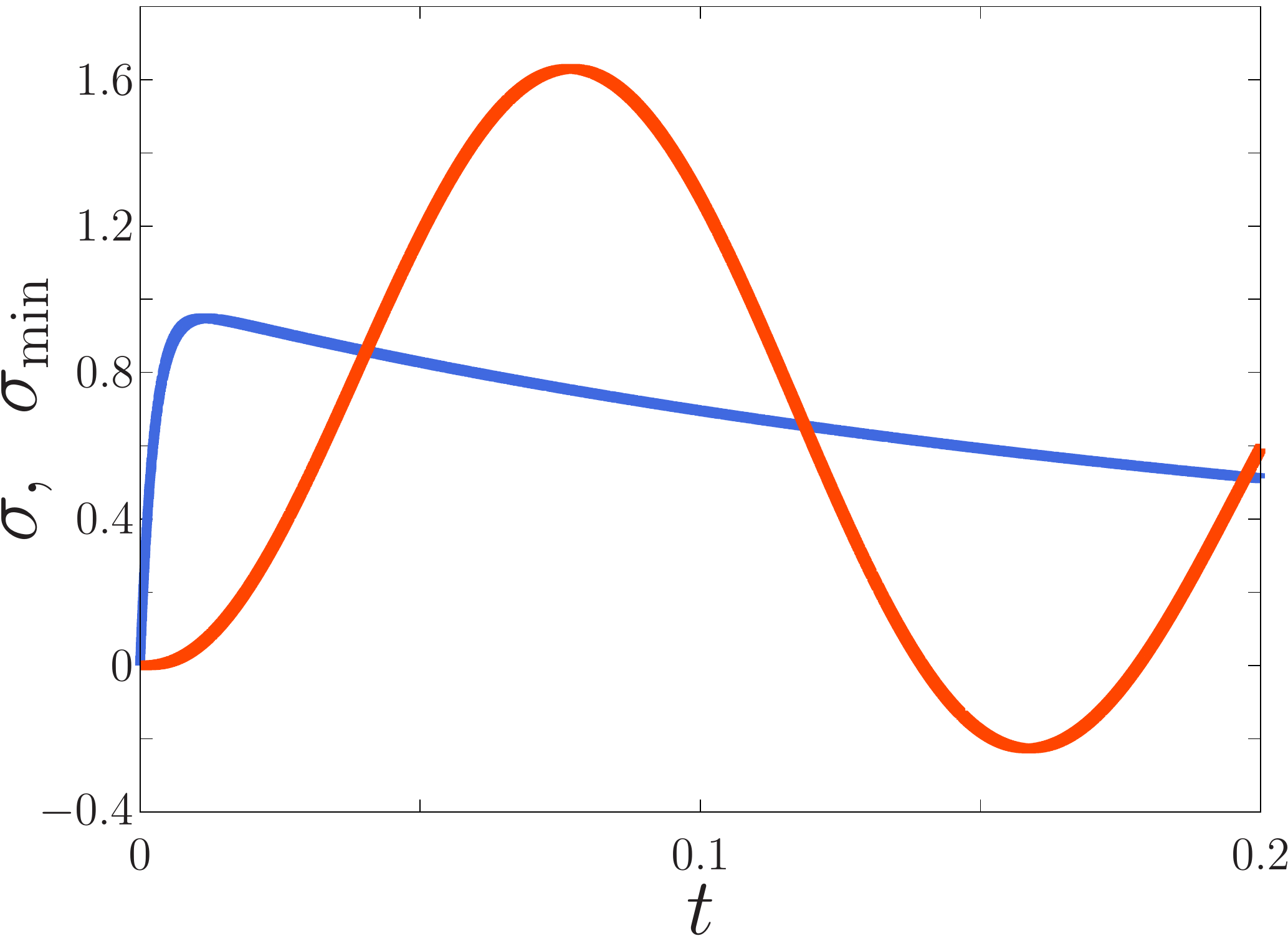}
	\end{minipage}
	\begin{minipage}{0.32\hsize}
		\centering
		\includegraphics[width=\linewidth]{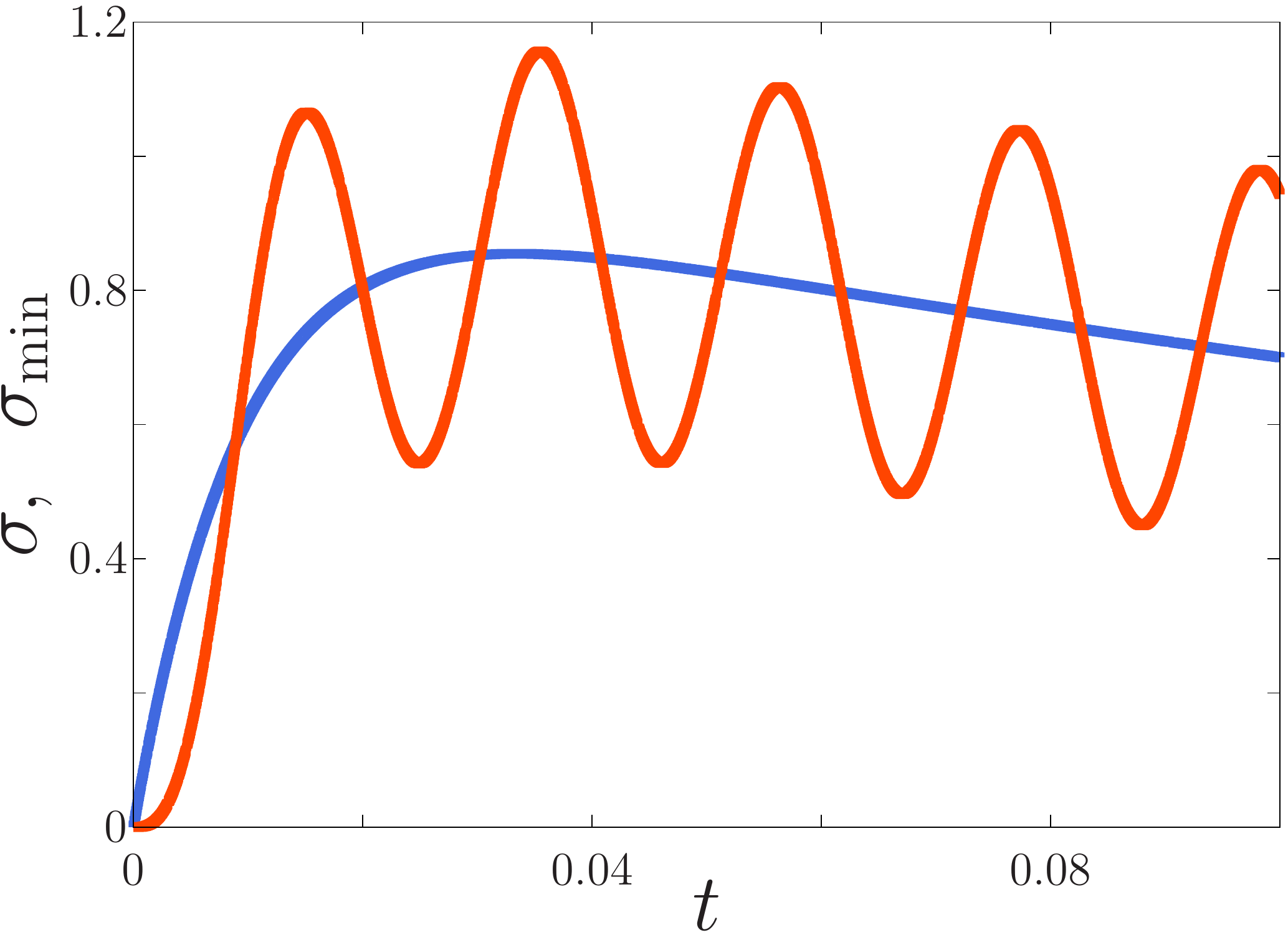}
	\end{minipage}
	\begin{minipage}{0.32\hsize}
		\centering
		\includegraphics[width=\linewidth]{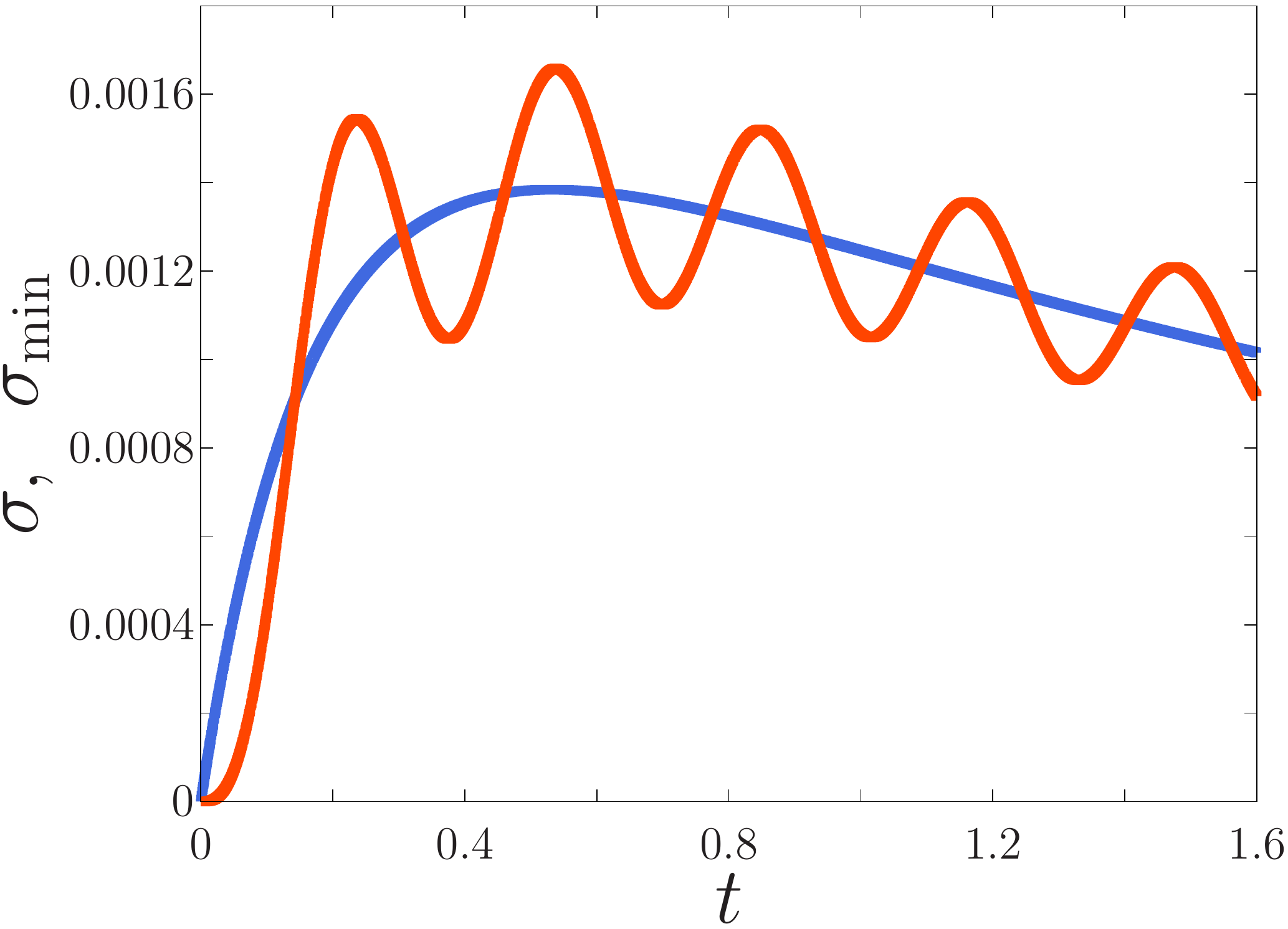}
	\end{minipage}
	\caption{
	\small Time evolutions of the modulus $\sigma(t)$ (\textcolor{red}{red}) and the modulus potential minimum $\sigma_\text{min}(t)$ (\textcolor{blue}{blue}). The left, central and right figure corresponds to the case of $(m_\sigma/H_\text{inf},\Gamma_\phi/H_\text{inf})=(40,400),~(300,100)~\text{and}~(20,0.01)$, respectively. The horizontal axis is $H_\text{inf} t$ with $t_0=0$. The vertical axis is $\sigma$ or $\sigma_\text{min}$ in unit of $(30/\pi^2g_*)(3M_P^2H_\text{inf}^4/Mm_\sigma^4)$. These figures are obtained by numerically solving Eqs.~\eqref{re_modulus}, \eqref{re_inflaton}, \eqref{re_radiation} and \eqref{re_friedmann}.
	}
	\label{fig:reheating}
\end{figure}
\begin{itemize}
\item $H_\text{inf}\ll m_\sigma\ll \Gamma_\phi$\\
In this case the reheating is instantaneous, and the modulus begins to oscillate after reheating is completed, as shown in the left figure of Fig.~\ref{fig:reheating}.
Since the cosmic expansion from $t_R$ to $t_\text{os}$ is negligible for $H_\text{inf}\ll m_\sigma$, we can take $T_\text{os}\simeq T_R$. From Eq.~\eqref{abundance} the modulus abundance for $t\geq t_\text{os}$ is obtained as
\begin{align}
	\frac{\rho_\sigma}{s}(t)&\simeq\frac{45}{4\pi^2g_*}\frac{T_R^5}{M^2m_\sigma^2},
	\label{abundance_1}
\end{align}
where we have used $3H_\text{inf}^2M_P^2=(\pi^2g_*/30)T_R^4$.

\item $H_\text{inf}\ll \Gamma_\phi\ll m_\sigma$\\
In this case the reheating is again instantaneous, but the modulus begins to oscillate before the reheating is completed, as shown in the central figure of Fig.~\ref{fig:reheating}. The energy density of the radiation at $t=t_\text{os}$ is roughly given by
\begin{align}
	\rho_\chi(t_\text{os})\simeq \Gamma_\phi\rho_\phi(t_0)\cdot (t_\text{os}-t_0)\simeq \frac{3\Gamma_\phi H_\text{inf}^2M_P^2}{m_\sigma}
	\simeq\frac{\pi^2 g_*}{30}T_R^4\frac{\Gamma_\phi}{m_\sigma},
\end{align}
where we have used $3H_\text{inf}^2 M_P^2=(\pi^2g_*/30)T_R^4$. 
Since $\rho_\chi(t_\text{os})=(\pi^2 g_*/30)T_\text{os}^4$, we find $T_\text{os}^4=T_R^4\cdot(\Gamma_\phi/m_\phi)$. From Eq.~\eqref{abundance} the modulus abundance for $t\geq t_R$ is obtained as
\begin{align}
	\frac{\rho_\sigma}{s}(t)
	\simeq \frac{45}{4\pi^2g_*}\frac{T_R^5}{M^2m_\sigma^2}\left(\frac{\Gamma_\phi}{m_\sigma}\right)^2.
	\label{abundance_2}
\end{align}

\item $\Gamma_\phi\ll H_\text{inf}\ll m_\sigma$\\
For $\Gamma_\phi\ll H_\text{inf}$, the temperature reaches its maximum value typically in about one Hubble time. Therefore in this case the modulus begins to oscillate before the temperature becomes maximum, as shown in the right figure of Fig.~\ref{fig:reheating}.
The energy density of the radiation at $t=t_\text{os}$ is roughly given by
\begin{align}
	\rho_\chi(t_\text{os})\simeq\frac{3\Gamma_\phi H_\text{inf}^2M_P^2}{m_\sigma}
	\simeq\frac{\pi^2 g_*}{30}T_R^4\frac{H_\text{inf}^2}{m_\sigma\Gamma_\phi},
\end{align}
where we have used $3\Gamma_\phi^2M_P^2=(\pi^2g_*/30)T_R^4$.
Since $\rho_\chi(t_\text{os})=(\pi^2 g_*/30)T_\text{os}^4$, we find $T_\text{os}^4=T_R^4\cdot(H_\text{inf}^2/m_\sigma\Gamma_\phi)$.
From Eq.~\eqref{abundance} the modulus abundance for $t\geq t_R$ is obtained as
\begin{align}
	\frac{\rho_\sigma}{s}(t)
	\simeq \frac{45}{4\pi^2g_*}\frac{T_R^5}{M^2m_\sigma^2}\left(\frac{H_\text{inf}}{m_\sigma}\right)^2.
	\label{abundance_3}
\end{align}
\end{itemize}

We have checked that Eqs.~\eqref{abundance_1}, \eqref{abundance_2} and \eqref{abundance_3} are in good agreement with the numerical calculation. Combining them, the abundance is summarized as
\begin{align}
	\frac{\rho_\sigma}{s} \simeq \frac{45}{4\pi^2g_*}\frac{T_R^5}{M^2m_\sigma^2} \epsilon
	\simeq 2\times 10^5\,{\rm GeV}\left( \frac{T_R}{10^{10}\,{\rm GeV}} \right)^5\left( \frac{1\,{\rm TeV}}{m_\sigma} \right)^2 \left( \frac{M_P}{M} \right)^2\left(\frac{106.75}{g_*}\right) \epsilon,
\end{align}
where
\begin{align}
	\epsilon\equiv {\rm max}\left[ {\rm min}\left[1,\frac{\Gamma_\phi^2}{m_\sigma^2}\right],~\frac{H_{\rm inf}^2}{m_\sigma^2}\right].
\end{align}
Note that $\epsilon \leq 1$ since we are assuming $H_{\rm inf} < m_\sigma$.
Though it is suppressed compared to the case of $H_\text{inf}\gtrsim m_\sigma$ given in Eq.~(\ref{abundance_usual}), it is not negligible and may have great impacts on cosmology in general.
However, one should take these results with a grain of salt. 
First, the instantaneous thermalization is a rather strong assumption, which may not be justified at the very beginning of the reheating.
Second, Eq.~(\ref{dotO/O}) formally diverges at $t\to t_0$ but it is an artifact of the approximation that the reheating started strictly at $t=t_0$ with a perturbative decay rate of the inflaton $\Gamma_\phi$ at the same time that the inflaton starts to behave as non-relativistic matter.
In reality, transition from inflation to the reheating regime is not sudden but smooth and the particle production rate may not be characterized just by perturbative decay rate. In the next section we perform more elaborated analyses taking account of the smooth transition and detailed particle production processes in a toy model.

\section{Moduli oscillation induced by (p)reheating}
\label{sec:example}

\subsection{Setup}
\label{sec:setup}

In this section, we analyze a toy model and estimate the modulus abundance coming from a rapid minimum shift during reheating, especially focusing on preheating, in order to clarify the argument in Sec.~\ref{sec:mechanism}.
As mentioned there, such a modulus production is likely to occur during reheating. However, the process of reheating is generally too complicated to track the whole history from the end of inflation to the complete thermalization. In order to avoid the complexity, we consider a simple system containing only three real scalars and concentrate on the stage of preheating. One of the scalars is the inflaton $\phi$. Another scalar plays a role of radiation interacting with the inflaton through a four-point interaction, which we denote by $\chi$. The other one is the modulus $\sigma$ linearly coupled to the radiation.\footnote{We assume that there is no direct coupling between the inflaton and the modulus. Otherwise, the modulus can be produced in a different way during preheating~\cite{Giudice:2001ep}.}
The action we adopt is
\begin{align}
	S = \int d^4x \sqrt{-g} \left(\frac{M_P^2}{2}R + \mathcal L_\phi + \mathcal L_\chi + \mathcal L_\sigma \right),
\end{align}
where $R$ denotes the Ricci scalar and each term will be explained below in detail.
We use the Friedmann-Robertson-Walker metric: $g_{\mu\nu}dx^\mu dx^\nu=a(\tau)^2(-d\tau^2+d\vec{x}^2)$, where $\tau$ is the conformal time and $a$ is the cosmic scale factor. Analytical and numerical estimations of the modulus abundance in the model are given in Sec.~\ref{sec:analytical} and Sec.~\ref{sec:numerical}, respectively.

\subsubsection{Inflaton sector}
\label{sec:inflaton_radiation}

We adopt the following action for the inflaton sector:
\begin{align}
	\mathcal L_\phi = -\frac{g^{\mu\nu}\partial_\mu\phi\partial_\nu\phi}{2\left(1-\phi^2/\Lambda^2\right)^{2}} - V(\phi),
\end{align}
with $\Lambda$ being a mass scale and $V(\phi)$ the inflaton potential.
We have taken the non-minimal inflaton kinetic term, which is known to induce the $\alpha\hf$attractor inflation~\cite{Kallosh:2013hoa,Ferrara:2013rsa,Kallosh:2013yoa,Galante:2014ifa}. We are interested in the case of $m_\sigma\gg H_\text{inf}$. 
For that purpose, the $\alpha\hf$attractor inflation is one of the favorable models since inflation energy scale can be arbitrarily lowered while keeping the consistency of the primordial curvature perturbation with the observation.
The canonical inflaton field $\varphi$ is defined by the following relation:
\begin{align}
	\phi=\Lambda\tanh\left(\frac{\varphi}{\Lambda}\right).
	\label{canonical}
\end{align}
Note that $\varphi\simeq\phi$ at around the origin. The inflaton field is regarded as a classical background filed, which induces $\chi$'s particle production.
The equation of motion for the inflaton reads
\begin{align}
	\varphi''+2\mathcal H\varphi'+a^2\frac{\partial V(\varphi)}{\partial\varphi} = 0,
	\label{eom_inflaton}
\end{align}
where the prime denotes the derivative with respect to $\tau$.
Here $\mathcal H\equiv a^{-1}(da/d\tau)$ is the conformal Hubble parameter, which is given by the Friedmann equation as
\begin{align}
	\mathcal H= \frac{1}{\sqrt{3}M_P}\sqrt{\frac{1}{2}\varphi'^2+a^2 V(\varphi)}.
	\label{friedmann}
\end{align}
In Eqs.~\eqref{eom_inflaton} and \eqref{friedmann}, we have neglected the contributions of $\chi$ and $\sigma$, which is justified as long as their energy densities are both subdominant to that of the inflaton.
The time evolution of the background fields can be obtained by solving these equations.

Let us move on to the inflaton potential $V(\phi)$.
We adopt the following potential:
\begin{align}
	V(\phi)=\frac{1}{2}m_\phi^2\phi^2,
	\label{potential_phi}
\end{align}
which becomes in terms of $\varphi$
\begin{align}
	V(\varphi) = \frac{1}{2}m_\phi^2\Lambda^2\tanh^2\left(\frac{\varphi}{\Lambda}\right).
	\label{potential_varphi}
\end{align}
Here, $m_\phi$ is the inflaton mass. Though the observed density perturbation, $\mathcal P_\xi\simeq 2.1\times 10^{-9}$~\cite{Aghanim:2018eyx}, almost fixes $m_\phi$ as $m_\phi\simeq 1.7\times 10^{13}\,\text{GeV}\,(50/N_e)$ with $N_e$ being the number of e-folding, the mass scale $\Lambda$ is arbitrary and related to the Hubble parameter during inflation as
\begin{align}
	H_\text{inf}\simeq 7.0\times 10^{12}\,\text{GeV}\left(\frac{\Lambda}{M_P}\right) \left(\frac{50}{N_e}\right).
\end{align}
Therefore, we can control the inflation scale $H_\text{inf}$ by varying $\Lambda$. The value of $\varphi$ at the end of inflation is given by
\begin{align}
	\varphi_\text{end}\simeq \frac{\Lambda}{2}\ln\left(\frac{4\sqrt{2}M_P}{\Lambda}\right).
	\label{phi_end}
\end{align}
The inflaton starts to oscillate around the potential minimum after the inflation ends. The oscillation is far from harmonic at the early stage especially for $\Lambda/M_P\ll1$, which makes the decay process of the inflaton complicated.

\subsubsection{Radiation sector}
\label{sec:radiation}

The radiation field $\chi$ is assumed to be a real scalar field and have the following simple form:
\begin{align}
	\mathcal L_\chi = -\frac{1}{2}g^{\mu\nu}\partial_\mu \chi \partial_\nu \chi - \frac{1}{2}g^2\phi^2\chi^2 - \frac{1}{2}\xi R \chi^2,
\end{align}
where $g$ and $\xi$ are coupling constants. Thus the radiation is assumed to couple with the inflaton via the four point interaction.
The interaction induces a time-dependent effective mass for $\chi$, which generally leads to $\chi$'s particle production.
We present the formulation of $\chi$'s particle production in our model, by following Ref.~\cite{Kofman:1997yn}. 

We define $\tilde{\chi}\equiv a\chi$, whose equation of motion is obtained as
\begin{align}
	\tilde{\chi}^{\prime\prime}-\partial_i^2\tilde{\chi}+m_{\chi,\text{eff}}^2\tilde{\chi} = 0,\quad
	m_{\chi,\text{eff}}^2\equiv a^2g^2\Lambda^2\tanh^2\left(\frac{\varphi}{\Lambda}\right)-(1-6\xi)\frac{a^{\prime\prime}}{a}.
	\label{eom_chi}
\end{align}
While $\varphi$ is the classical background field, $\tilde{\chi}$ is quantized as
\begin{align}
	\tilde{\chi}(\tau,\vec{x})=\int\frac{d^3k}{(2\pi)^3}[a_{\vec{k}}\chi_k(\tau)+a^\dagger_{-\vec{k}}\chi_k^*(\tau)]e^{i\vec{k}\cdot\vec{x}}.
	\label{chi_quantize}
\end{align}
From Eq.~\eqref{eom_chi}, the mode function $\chi_k$ satisfies the following equation:
\begin{align}
	\chi_k''+\omega_k^2\chi_k=0,\quad
	\omega_k^2\equiv k^2+m_{\chi,\text{eff}}^2.
	\label{eom_mode}
\end{align}
$a_{\vec{k}}$ and $a^\dagger_{\vec{k}}$ are the creation and annihilation operators of $\chi$ particles with the comoving wave number $\vec{k}$, which satisfy
\begin{align}
	[a_{\vec{k}},a^\dagger_{\vec{k}'}]=(2\pi)^3\delta(\vec{k}-\vec{k}'),~~~[a_{\vec{k}},a_{\vec{k}'}]=[a^\dagger_{\vec{k}},a^\dagger_{\vec{k}'}]=0.
\end{align}
Then the canonical commutation relation $[\tilde{\chi}(\tau,\vec{x}),\tilde{\chi}'(\tau,\vec{x}')]=i\delta^3(\vec{x}-\vec{x}')$ leads to
\begin{align}
	\chi_k\chi_k'^*-\chi_k^*\chi_k'=i,
	\label{normalization}
\end{align}
which fixes the normalization of $\tilde{\chi}$.
With $\tau_i$ being the initial time, we take the following initial condition:
\begin{align}
	\chi_k(\tau\rightarrow\tau_i)\simeq v_k(\tau),~~~\chi_k'(\tau\rightarrow\tau_i)\simeq-i\omega_k(\tau) v_k(\tau),
	\label{chi_initial}
\end{align}
where
\begin{align}
	v_k(\tau)=\frac{1}{\sqrt{2\omega_k(\tau)}}\exp\left(-i\int_{\tau_i}^\tau\omega_k(\tau') d\tau'\right).
	\label{v_k}
\end{align}
Note that the vacuum state satisfies $a_{\vec{k}}|0\rangle=0$ at the initial time $\tau_i$.

The solution of Eqs.~\eqref{eom_mode}, \eqref{normalization} and \eqref{chi_initial} can be expressed as
\begin{align}
	\chi_k(\tau)=\alpha_k(\tau) v_k(\tau)+\beta_k(\tau) v_k^*(\tau),
	\label{expansion}
\end{align}
where $\alpha_k(\tau)$ and $\beta_k(\tau)$ are given by
 \begin{align}
	\frac{d\alpha_k}{d\tau} = \frac{\beta_k}{2\omega_k}\frac{d\omega_k}{d\tau}\exp\left(2i\int_{\tau_i}^\tau\omega_kd\tau\right),~~~
	\frac{d\beta_k}{d\tau} = \frac{\alpha_k}{2\omega_k}\frac{d\omega_k}{d\tau}\exp\left(-2i\int_{\tau_i}^\tau\omega_kd\tau\right),
	\label{diffeq_bogo}
\end{align}
with initial conditions $\alpha_k(\tau_i)=1$ and $\beta_k(\tau_i)=0$.
The complex variables $\alpha_k(\tau)$ and $\beta_k(\tau)$ are the so-called Bogolyubov coefficients,
and satisfy $|\alpha_k(\tau)|^2-|\beta_k(\tau)|^2=1$.
We adopt the following definition of the occupation number:
\begin{align}
	n_{\chi,k}(\tau)&\equiv\frac{1}{2\omega_k}\left[ |\chi_k'|^2+\omega_k^2|\chi_k|^2 \right]-\frac{1}{2}=|\beta_k(\tau)|^2,
	\label{occupation}
\end{align}
where we have subtracted the vacuum fluctuation.
Then the number and energy densities of $\chi$ are expressed as
\begin{align}
	n_\chi(\tau)&\equiv a(\tau)^{-3}\int\frac{d^3k}{(2\pi)^3}n_{\chi,k}(\tau)= a(\tau)^{-3}\int\frac{d^3k}{(2\pi)^3}|\beta_k(\tau)|^2,\label{chi_number}\\
	\rho_\chi(\tau)&\equiv a(\tau)^{-4}\int\frac{d^3k}{(2\pi)^3}\omega_k(\tau)n_{\chi,k}(\tau)= a(\tau)^{-4}\int\frac{d^3k}{(2\pi)^3}\omega_k(\tau)|\beta_k(\tau)|^2.\label{chi_energy}
\end{align}
The particle production is characterized by $\beta_k(\tau)\neq 0$.
The evolution of $\beta_k(\tau)$ is obtained by solving Eq.~\eqref{diffeq_bogo} with the background values satisfying Eqs.~\eqref{eom_inflaton} and \eqref{friedmann}.

Here, we comment on the reheating in this model.
At the end of the inflation, the inflaton condensate dominates the energy density of the universe. Then, the inflaton decay proceeds through particle production processes coming from the four-point interaction $\sim g^2 \phi^2\chi^2$.
In general, however, reheating may not be completed only with this interaction since the inflaton sector has a $Z_2$ symmetry that makes the inflaton stable against the decay~\cite{Mukaida:2013xxa}.
Therefore, the existence of some perturbative decay and thermalization process as well as necessary particle contents are implicitly assumed in this model, although it does not affect the following discussion at all.

\subsubsection{Modulus sector}
\label{sec:modulus}

We adopt the following Lagrangian for the modulus $\sigma$:\footnote{We ignore the back reaction of the last term on $\chi$.
Such an approximation is typically valid as far as $|\sigma| \lesssim M$, and it is the case in the discussion in Sec.~\ref{sec:analytical} and Sec.~\ref{sec:numerical}.}
\begin{align}
	\mathcal L_\sigma = -\frac{1}{2}g^{\mu\nu}\partial_\mu\sigma\partial_\nu\sigma-\frac{1}{2}m_\sigma^2\sigma^2+\frac{\sigma}{M}\mathcal O,\quad
	\mathcal O=\chi\Box\chi,
	\label{action_sigma}
\end{align}
where $m_\sigma$ is the modulus mass, $M$ is a mass scale,
and $\Box$ is the d'Alembertian $a^{-2}(-\partial_\tau^2+\partial_i^2)$.
If there is no particle production of $\chi$, the linear term has no effects on the dynamics of $\sigma$. Otherwise, it deviates the minimum of the modulus potential from the origin. If the time scale of the shift is shorter than $m_\sigma^{-1}$, the modulus cannot track the minimum adiabatically and the oscillation can be induced as discussed in Sec.~\ref{sec:mechanism}.
Note that, if the modulus mass is larger than the inflaton mass, the oscillation may be significantly suppressed.
In order to make the comparison between the analytic estimation and the numerical simulation clear, we hereafter focus on the following region:
\begin{align}
	H_\text{inf}\ll m_\sigma\ll m_\phi.
	\label{region}
\end{align}

Let us derive an expression for the modulus abundance in this model.
From Eq.~\eqref{action_sigma}, the equation of motion for the modulus reads
\begin{align}
	\sigma^{\prime\prime}+2\mathcal H\sigma^\prime+a^2\left[ m_\sigma^2\sigma-\langle\mathcal O\rangle/M \right] = 0.
	\label{eom_modulus}
\end{align}
We first evaluate the vacuum expectation value of $\mathcal O$, which is given by
\begin{align}
	\langle\mathcal O\rangle=
	\langle\chi\Box\chi\rangle
	=\frac{1}{a^4}
	\langle\mathcal C\tilde{\chi}^2+\mathcal H (\tilde{\chi}\tilde{\chi}^\prime+\tilde{\chi}^\prime\tilde{\chi})\rangle,
\end{align}
where
\begin{align}
	\mathcal C
	\equiv m_{\chi,\text{eff}}^2+\frac{a^{\prime\prime}}{a}-2\mathcal H^2
	=a^2g^2\Lambda^2\tanh^2\left(\frac{\varphi}{\Lambda}\right)+6\xi \frac{a^{\prime\prime}}{a}-2\mathcal H^2.
	\label{parameter_c}
\end{align}
As long as $m_\sigma\ll m_\phi$, we can regard $\tanh^2(\varphi/\Lambda)$ in $\mathcal C$ as $\tanh^2(\Phi/\Lambda)$, where $\Phi(t)$ is the oscillation amplitude of $\varphi$.
In the last expression of Eq.~\eqref{parameter_c}, the last two terms are suppressed compared to the first term by a factor of $\mathcal H^2/\mathcal C\simeq H^2/g^2\Lambda^2$ which is typically much smaller than unity.
Therefore we can take $\mathcal C\simeq a^2g^2\Lambda^2\tanh^2(\Phi/\Lambda)$.
Then $\langle\mathcal O\rangle$ is approximated as
\begin{align}
	\langle\mathcal O\rangle\simeq \frac{\mathcal C\langle\tilde{\chi}^2\rangle}{a^4}
	\simeq g^2\Lambda^2\tanh^2\left(\frac{\Phi}{\Lambda}\right)\langle\chi^2\rangle,
	\label{O_approximated}
\end{align}
where we have neglected $\mathcal H\langle\tilde{\chi}\tilde{\chi}^\prime+\tilde{\chi}^\prime\tilde{\chi}\rangle$ since it is suppressed compared to $\mathcal C\langle\tilde{\chi}^2\rangle$ again by a factor of $\mathcal H^2/\mathcal C\simeq H^2/g^2\Lambda^2$.
From Eq.~\eqref{condition}, the condition that the modulus oscillation occurs is written as
\begin{align}
	m_\sigma\ll\left|\frac{1}{\langle\chi^2\rangle}\frac{d\langle\chi^2\rangle}{dt}\right|.
	\label{sigma_condition}
\end{align}
Eqs.~\eqref{minimum} and \eqref{O_approximated} lead to the initial oscillation amplitude as
\begin{align}
	\hat{\sigma}(t_\text{os})\simeq\sigma_\text{min}(t_\text{os})
	\simeq\frac{\langle\mathcal O\rangle(t_\text{os})}{Mm_\sigma^2}
	\simeq\frac{g^2\Lambda^2\tanh^2(\Phi_\text{os}/\Lambda)}{Mm_\sigma^2}\langle\chi^2\rangle(t_\text{os}),
\end{align}
where $\Phi_\text{os}\equiv\Phi(t=t_\text{os})$.
The initial energy density of the modulus oscillation is calculated from Eqs.~\eqref{energy} and \eqref{O_approximated} as
\begin{align}
	\rho_\sigma(t_\text{os})\simeq\frac{1}{2}m_\sigma^2\hat{\sigma}^2(t_\text{os})
	\simeq\frac{g^4\Lambda^4\tanh^4(\Phi_\text{os}/\Lambda)}{2M^2m_\sigma^2}[\langle\chi^2\rangle(t_\text{os})]^2.
	\label{sigma_energy}
\end{align}
The energy density over the entropy density is time-independent for $t\geq t_R$ and calculated from Eqs.~\eqref{abundance} and \eqref{O_approximated} as
\begin{align}
	\frac{\rho_\sigma}{s}
	\simeq\frac{3T_R}{4}\frac{g^4\Lambda^2\tanh^2(\Phi_\text{os}/\Lambda)}{M^2m_\sigma^2m_\phi^2}[\langle\chi^2\rangle(t_\text{os})]^2,
	\label{matter_dominated}
\end{align}
where we have used $3H^2_\text{inf}M_P^2\simeq m_\phi^2\Lambda^2\tanh^2(\Phi_\text{os}/\Lambda)/2$.

One can see that in the present case $\langle\chi^2\rangle$ determines whether the modulus oscillation is induced or not as well as how much the abundance is.
By using Eqs.~\eqref{chi_quantize}, \eqref{expansion} and \eqref{diffeq_bogo}, $\langle\chi^2\rangle$ can be expressed in terms of the Bogolyubov coefficients as
\begin{align}
	\langle\chi^2\rangle=\frac{1}{a^2}\int\frac{d^3 k}{(2\pi)^3}\frac{1}{\omega_k}(|\beta_k|^2+\text{Re}\mathcal A),
	\label{chi_fluctuation}
\end{align}
where
\begin{align}
	\mathcal A(\tau)\equiv \alpha_k(\tau)\beta_k^*(\tau) \exp\left(-2i\int_{\tau_i}^\tau\omega_k(\tau') d\tau'\right).
\end{align}
We have subtracted the divergent constant term in the momentum integral by renormalization.\footnote{In this case, $\langle\chi^2\rangle$ vanishes for $\beta_k=0$, which is consistent with the fact that there is no $\chi$'s particle production at the initial time.}
In Sec.~\ref{sec:evaluation}, we evaluate $\langle\chi^2\rangle$ during preheating in the $\alpha\hf$attractor inflation.

\subsection{Analytical estimation}
\label{sec:analytical}

In this subsection, we discuss the modulus production in our toy model analytically.
First, we briefly explain the preheating in the $\alpha\hf$attractor inflation, which has been discussed in Refs.~\cite{Krajewski:2018moi,Iarygina:2018kee} in a different setup. Next, we give an analytical expression for the fluctuation of $\chi$ as well as the number density of $\chi$ in the broad resonance regime.
Finally, we introduce a special value of $g$ which is the main target of the numerical calculation in the next subsection.

\subsubsection{Preheating in the $\alpha\hf$attractor inflation}
\label{sec:preheating}

In the early stage of reheating, non-perturbative effects can play an important role, during which the number density of the produced particle grows exponentially via the so-called parametric resonance. Such a period is known as preheating. In the following, we give a qualitative discussion of the preheating in our setup and estimate the typical momentum region enhanced in the broad resonance regime.

The preheating in the present model is a little complicated because of the inflation model. For $\Lambda\gtrsim M_P$, the amplitude of the inflaton oscillation $\Phi(t)$ is smaller than $\Lambda$ after inflation because $\varphi_\text{end}\lesssim\Lambda$ from Eq.~\eqref{phi_end}.
Then the potential \eqref{potential_varphi} can be regarded as a quadratic one, and the preheating is almost the same as in the case of the chaotic inflation~\cite{Kofman:1997yn}.
In that case, the parametric resonance is characterized by the following parameter: $q(t)\equiv g^2\Phi(t)^2/4m_\phi^2$.
$q\gg 1$ ($q\ll 1$) corresponds to the so-called broad (narrow) resonance regime.
For $\Lambda\lesssim M_P$, however, the non-quadratic nature of the inflaton potential becomes crucial.
Actually, we are interested in the low scale inflation, i.e. $\Lambda\ll M_P$, and hence the situation of our interest is far from the quadratic case.
Nevertheless, the following qualitative discussion may be useful to estimate the efficiency of preheating.

\begin{figure}[t]
	\begin{minipage}{0.5\hsize}
		\centering
		\includegraphics[width=\linewidth]{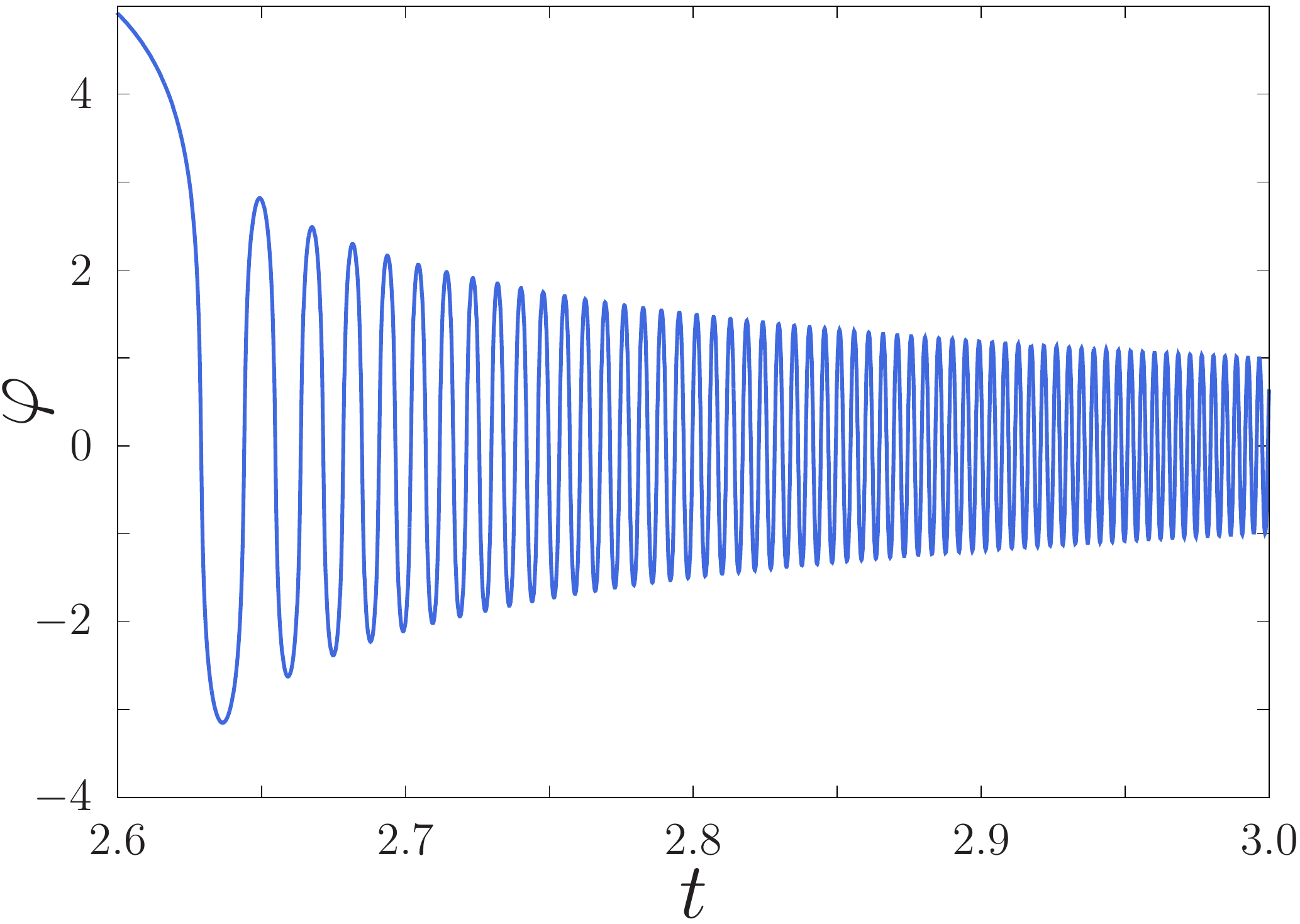}
	\end{minipage}
	\begin{minipage}{0.5\hsize}
		\centering
		\includegraphics[width=\linewidth]{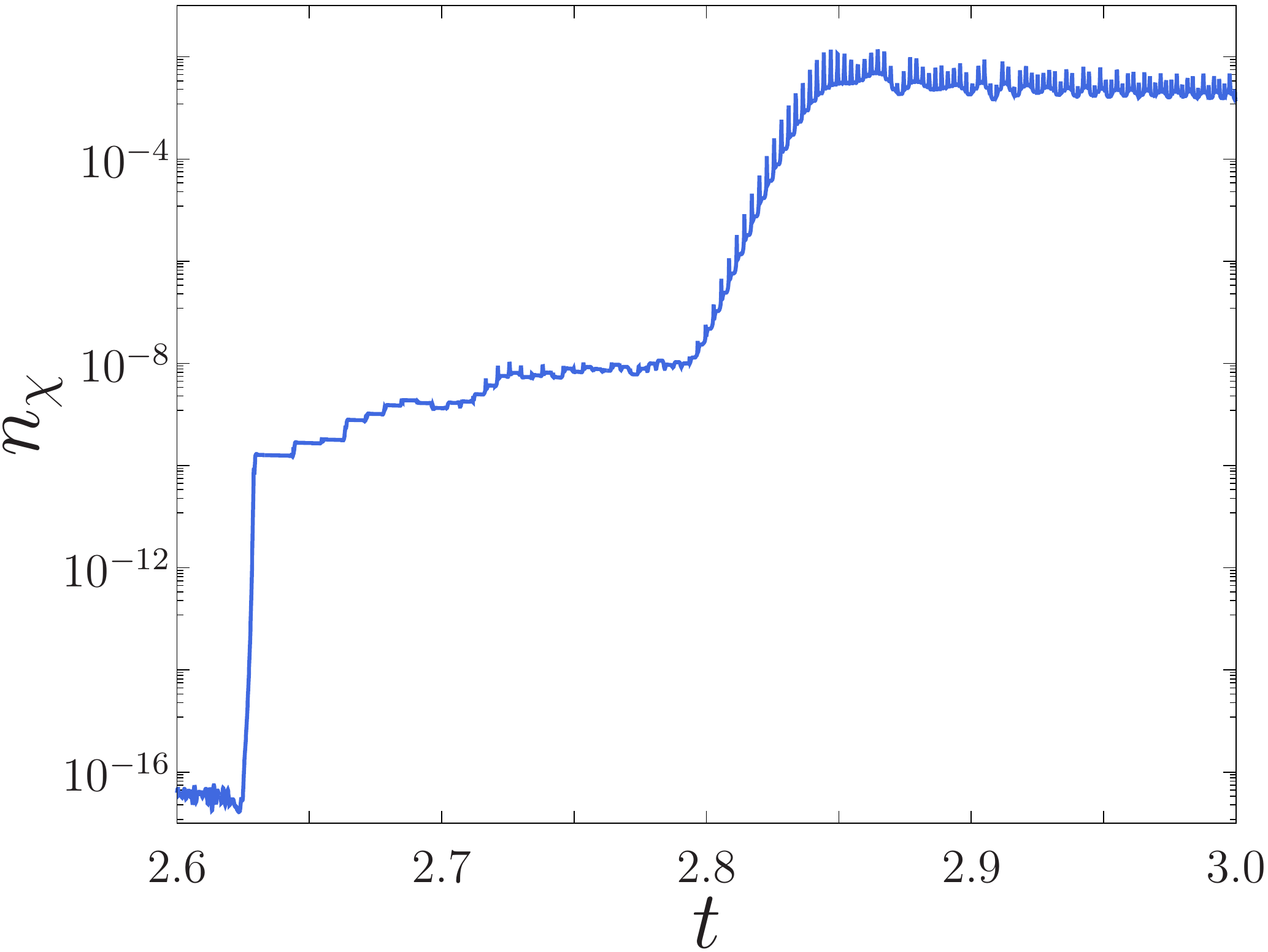}
	\end{minipage}
	\caption{
	\small Time evolutions of the inflaton position $\varphi(t)$ (left) and the number density of the radiation $n_\chi(t)$ (right) for $\Lambda/M_P=10^{-3}$. In the right figure, we take $g=4.1\times10^{-3}$. The horizontal axis is $H_\text{inf} t$ with the initial time $t=0$. The vertical axis of the left (right) figure is $\varphi/\Lambda$ ($n_\chi/\Lambda^3$). 
	}
	\label{fig:oscillation}
\end{figure}

From Eq.~\eqref{eom_mode}, the equation of motion for the mode function $\chi_k$ is given by
\begin{align}
	\left[ \frac{d^2}{d\tau^2}+k^2+m_{\chi,\text{eff}}^2(\tau)\right]\chi_k(\tau)=0.
	\label{mathieu_1}
\end{align}
If the expansion of the universe can be neglected, the equation is reduced to
\begin{align}
	\left[ \frac{d^2}{dt^2}+\frac{k^2}{a^2}+g^2\Lambda^2\tanh^2\left(\frac{\varphi(t)}{\Lambda}\right) \right]\chi_k(t)=0.
	\label{mathieu_2}
\end{align}
Due to the periodicity, there can be a exponentially growing solution, depending on the parameters.
If the cosmic expansion is included, it changes with time whether the system is in the stability or instability region. In that case, the resonance proceeds in a stochastic manner (see Fig.~\ref{fig:oscillation}).

Let us take a look at the growth of the occupation number. From Eqs.~\eqref{eom_mode} and \eqref{occupation}, we obtain
\begin{align}
	n_{\chi,k}'\sim \mathcal O\left(\frac{\omega_k'}{\omega_k^2}\right)\omega_k n_{\chi,k},
\end{align}
where we have used WKB approximation and dropped the subdominant terms in $\mathcal H^2/\omega_k^2$ expansion.
It is known that the growth of $n_{\chi,k}$ is inefficient as long as the following adiabaticity condition is satisfied:\footnote{Even if Eq.~\eqref{adiabaticity} is satisfied, large $n_{\chi,k}$ and long duration may lead to a sizable enhancement, which corresponds to the narrow resonance~\cite{Dolgov:1989us,PhysRevD.42.2491,Shtanov:1994ce}.}
\begin{align}
	\left|\frac{\omega_k'}{\omega_k^2}\right|\ll 1.
	\label{adiabaticity}
\end{align}
On the other hand, the violation of the above condition leads to an exponential enhancement of $n_{\chi,k}$. Such an enhancement corresponds to the broad resonance. 
In general Eq.~\eqref{adiabaticity} is likely to be violated when the inflaton is at around the origin. Therefore, in the broad resonance regime, the amplification occurs in a short period when the inflaton crosses the origin.

Let us estimate the typical momentum region enhanced in the broad resonance regime.
In order to evaluate the left hand side of Eq.~\eqref{adiabaticity}, we need $\dot{\varphi}(t)$ which can be found from the following energy conservation:
\begin{align}
	\frac{1}{2}m_\phi^2\Lambda^2\tanh^2\left(\frac{\Phi(t)}{\Lambda}\right)\simeq \frac{1}{2}m_\phi^2\Lambda^2\tanh^2\left(\frac{\varphi(t)}{\Lambda}\right)+\frac{1}{2}\dot{\varphi}^2(t).
	\label{energy_conservation}
\end{align}
We can take $\tanh^2(\Phi(t)/\Lambda)\simeq 1$ because $\Phi(t)\gtrsim\Lambda$ during the broad resonance in the parameter region of our interest. (See Table~\ref{tab:value}.)
In this case, Eq.~\eqref{energy_conservation} leads to
\begin{align}
	|\dot{\varphi}(t)|\simeq m_\phi\Lambda\sqrt{1-\tanh^2(\varphi/\Lambda)}=m_\phi\Lambda\cosh^{-1}(\varphi/\Lambda).
\end{align}
Then the left hand side of Eq.~\eqref{adiabaticity} is calculated as
\begin{align}
	\left|\frac{\omega_k'}{\omega_k^2}\right|&= \left|\frac{g^2\dot{\varphi}\Lambda\tanh(\varphi/\Lambda)\cosh^{-2}(\varphi/\Lambda)}{[k^2/a^2+g^2\Lambda^2\tanh^2(\varphi/\Lambda)]^{3/2}}\right|\\
	&\simeq\frac{m_\phi}{g\Lambda}\cdot\frac{z(1-z^2)^{3/2}}{(A^2+z^2)^{3/2}},
	\label{adiabaticity_function}
\end{align}
where $z\equiv\tanh(\varphi/\Lambda)$ ($|z|<1$) and $A\equiv (k/a)/(g\Lambda)$.
Now $|\omega_k'/\omega_k^2|$ is a function of $z$ and $A$.
For $A\ll 1$, $|\omega_k'/\omega_k^2|$ becomes maximum at $z\simeq A/\sqrt{2}$, and the maximum value decreases as $\sim (m_\phi/g\Lambda) A^{-2}$ with respect to $A$. 
For $A\gg 1$, $|\omega_k'/\omega_k^2|$ becomes maximum at $z\simeq 1/2$, and the maximum value decreases more rapidly as $\sim(m_\phi/g\Lambda) A^{-3}$ with respect to $A$.
For the parameter region of our interest, which will be discussed in Sec.~\ref{sec:target}, $m_\phi/g\Lambda$ is comparable to or a little larger than unity. (See Table~\ref{tab:value}.)
Therefore it is reasonable to suppose that the momentum region,
\begin{align}
	p=k/a\ll g\Lambda\left(\frac{m_\phi}{g\Lambda}\right)^{1/3} \equiv p_* (\equiv k_*/a),
\end{align}
is significantly enhanced in the broad resonance regime.
Although this is just a rough estimation using the adiabaticity condition (\ref{adiabaticity}) and we may require more sophisticated analyses to rigorously discuss the momentum distribution of $\chi$, we have checked that this estimation is compatible with the numerical results at least in the parameter region of our interest. 

\begin{figure}[t]
	\begin{minipage}{0.5\hsize}
		\centering
		\includegraphics[width=\linewidth]{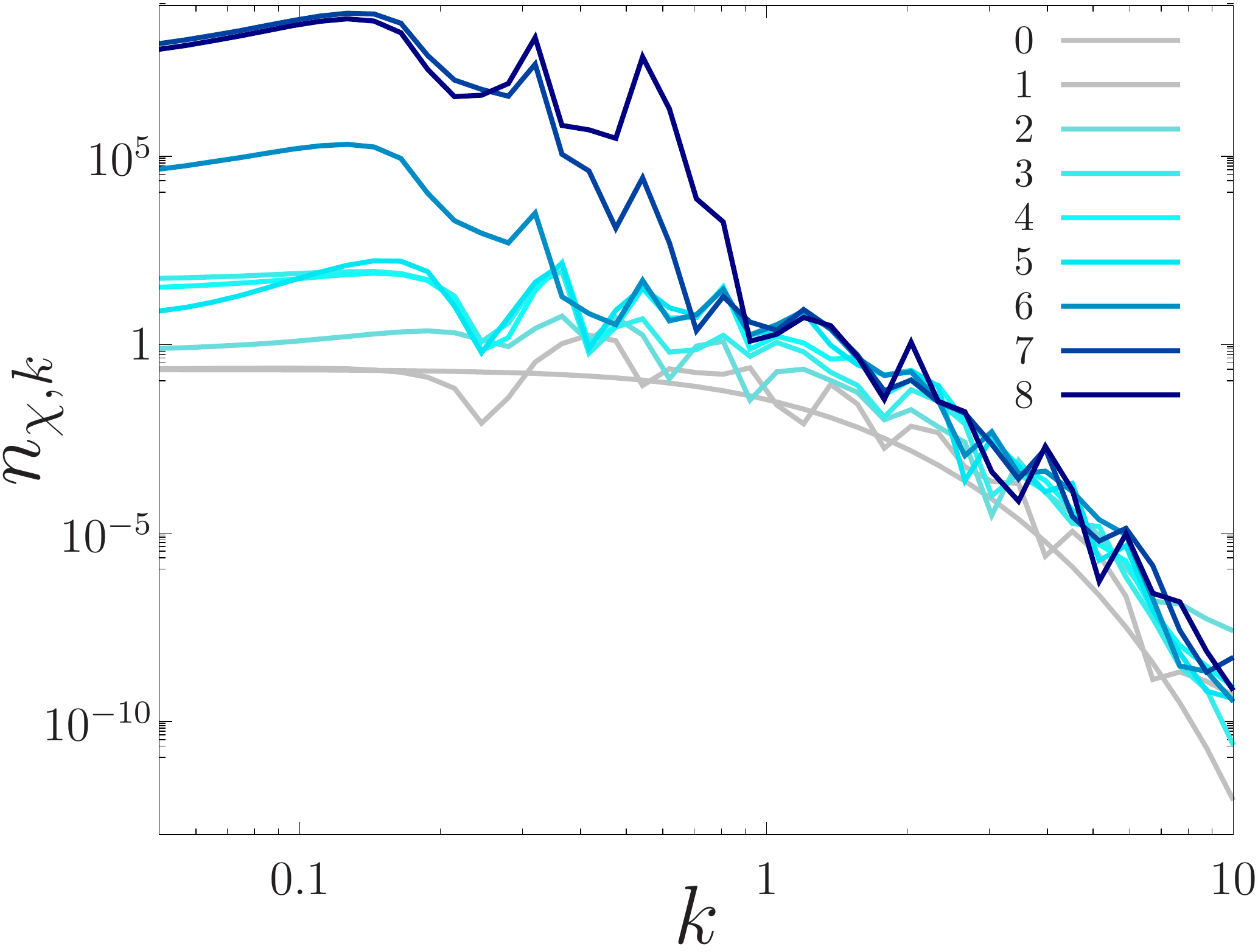}
	\end{minipage}
	\begin{minipage}{0.5\hsize}
		\centering
		\includegraphics[width=\linewidth]{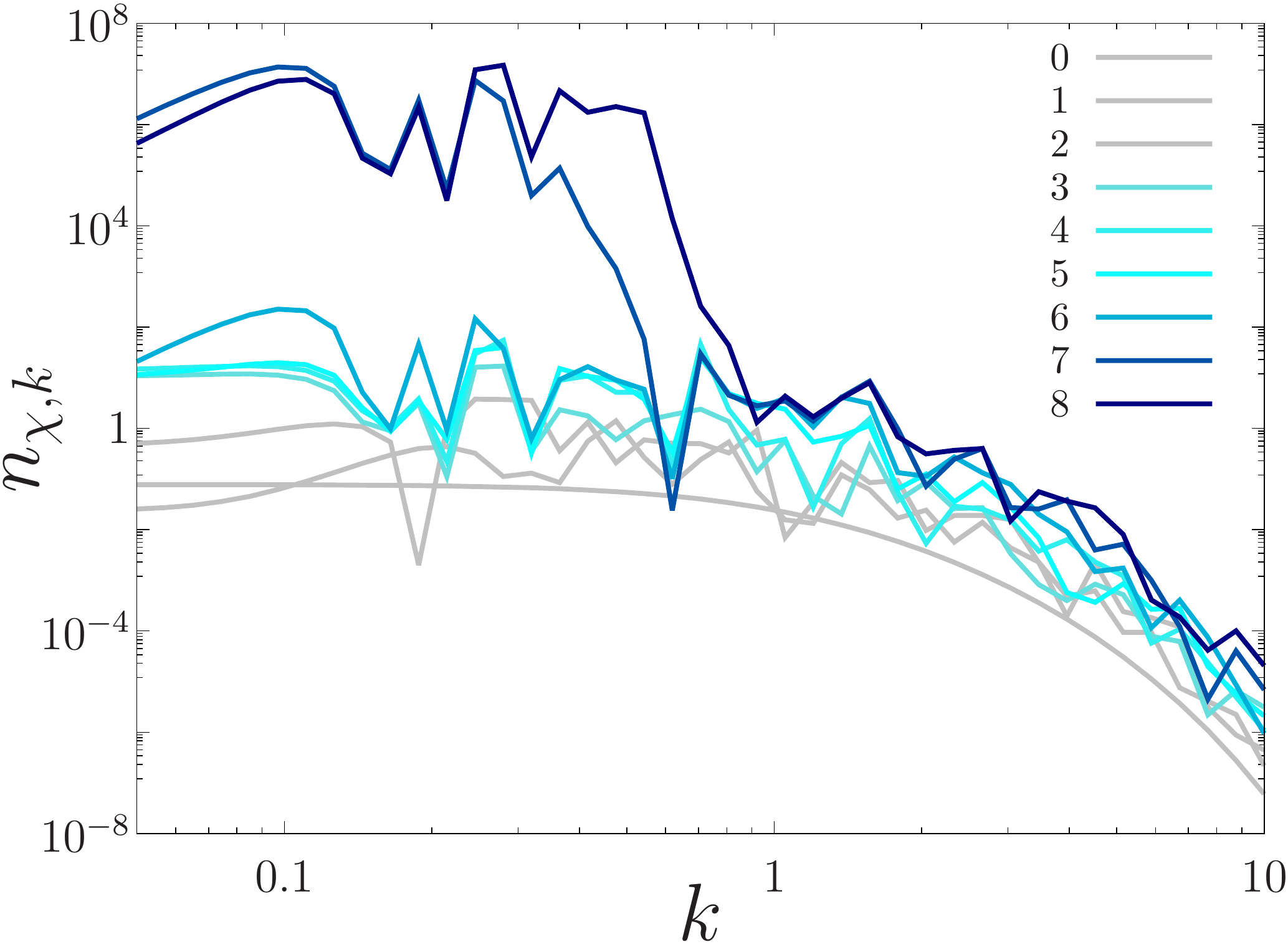}
	\end{minipage}
	\caption{
	\small Time evolution of $\chi$'s momentum distribution during preheating. The horizontal axis is the comoving wave number in unit of $ag\Lambda$ with $a$ evaluated at the first inflaton zero crossing. The vertical axis is the occupation number $n_{\chi,k}$. The left and right figures correspond to the case of $(\Lambda/M_P,g)=(10^{-3},4.1\times10^{-3})~\text{and}~(10^{-4},1.8\times10^{-2})$, respectively.
	The lines labeled by $0$ to $8$ correspond to the occupation number at different time slices, which are taken at regular intervals from the first zero crossing of the inflaton to the end of the preheating.
		}
	\label{fig:distribution}
\end{figure}

In Fig.~\ref{fig:oscillation} and Fig.~\ref{fig:distribution}, we show some examples of numerical results. Here and hereafter, we take $\xi=0$ and $N_e=50$ for concreteness.\footnote{The results do not depend much on the choice unless $|\xi|$ is very large~\cite{Bassett:1997az,Tsujikawa:1999jh}.}
Fig.~\ref{fig:oscillation} shows the time evolutions of the inflaton position $\varphi(t)$ (left) and the number density of the radiation $n_\chi(t)$ (right) for $\Lambda/M_P=10^{-3}$.
In the left figure, we start the calculation from $t=0$ with the initial condition $\varphi(0)=2\varphi_\text{end}$. The corresponding initial velocity is derived from Eqs.~\eqref{eom_inflaton} and \eqref{friedmann} with slow-roll approximation $\varphi''(0)=0$. On the other hand, in the right figure, we take $g=4.1\times10^{-3}$ and start the calculation a little earlier than the first inflaton zero crossing with the initial condition $\alpha_k=1$ and $\beta_k=0$ for all $k$. The result is not sensitive to the choice of the initial point as long as it is placed before the first inflaton zero crossing. In the right figure, one can see that the growth rate of $n_\chi$ changes at every inflaton zero crossing. Even in this stochastic evolution, a relatively significant amplification can be seen for $2.8\lesssim H_\text{inf} t\lesssim 2.85$, which we call the broad resonance in this paper.

Fig.~\ref{fig:distribution} shows the time evolution of $\chi$'s momentum distribution during preheating. 
The horizontal axis is the comoving wave number in unit of $ag\Lambda$ with $a$ evaluated at the first inflaton zero crossing, and the vertical axis is the occupation number $n_{\chi,k}$. 
The left and right figures correspond to the case of $(\Lambda/M_P,g)=(10^{-3},4.1\times10^{-3})~\text{and}~(10^{-4},1.8\times10^{-2})$, respectively.
The lines labeled by 0 to 8 correspond to the occupation number $n_{\chi,k}(t_i)$ at different time slices $t_i$, which are taken at regular intervals from the first zero crossing of the inflaton to the end of the preheating.
In both figures, comoving wave numbers smaller than $\sim ag\Lambda$ are enhanced more significantly than any other regions.

\subsubsection{Evaluation of $\langle\chi^2\rangle$ during preheating}
\label{sec:evaluation}

Let us calculate the fluctuation of $\chi$ which is the main ingredient of the analytical expressions for the modulus production. We apply the discussion in Sec.~\ref{sec:modulus} to the case of preheating in the $\alpha\hf$attractor inflation.

If the broad resonance lasts long enough, the occupation number with momentum smaller than $p_*(\sim g\Lambda)$ is significantly amplified. As a result, noting $|\alpha_k|^2-|\beta_k|^2=1$, we can use an approximation $|\alpha_k|\simeq |\beta_k|\gg 1$ for $k\ll k_*(\sim ag\Lambda)$. Then in the momentum integral of Eq.~\eqref{chi_fluctuation} the region around $\sim k_*$ may give the dominant contribution, and $\langle\chi^2\rangle$ may be approximated as
\begin{align}
	\langle\chi^2\rangle&\simeq\frac{1}{a^2}\int\frac{d^3 k}{(2\pi)^3}\frac{|\beta_k|^2}{\omega_k}\left[ 1+\cos\left(2\int\omega_kd\tau-\text{arg}[\alpha_k\beta^*_k]\right)\right]\label{fluctuation_preheating_first}\\
	&\sim\frac{n_\chi}{g\Lambda}\left[ 1+\cos\left(2g\Lambda\int\tanh\left(\frac{\varphi(t)}{\Lambda}\right)dt+\vartheta\right)\right],
	\label{fluctuation_preheating}
\end{align}
where we have used Eq.~\eqref{chi_number} and an approximation $\omega_{k_*}\simeq m_{\chi,\text{eff}}\sim ag\Lambda$.
Here $\vartheta$ is a numerical factor coming from the contribution of $\text{arg}[\alpha_k\beta^*_k]$.
The second term in the last expression is a rapidly oscillating function whose oscillation time scale is $\sim g\Lambda$. As far as the modulus dynamics with the mass smaller than $\sim g\Lambda$ is concerned, which is often the case for the region \eqref{region}, the term can be safely neglected.

Now that $\langle\chi^2\rangle$ is related to $n_\chi$, we give a formal expression for $n_\chi$ in the broad resonance regime.
The amplification of the occupation number can be described as $n_{\chi,k}(t_\text{after})=e^{2\pi\mu_k}n_{\chi,k}(t_\text{before})$ where $t_\text{before}$ ($t_\text{after}$) is the time just before (after) some inflaton zero crossing.
We assume that in the broad resonance regime the growth factor $\mu_k$ has a peak at $\sim k_*$ and the second derivative at the peak is given by $\partial_k^2\mu_k|_{k=k_*}\sim2\mu_{k_*}/k_*^2$.
In this case, $n_\chi$ after a number of inflaton oscillations is calculated as~\cite{Kofman:1994rk,Kofman:1997yn}
\begin{align}
	n_\chi(t)&=\frac{1}{a^3(t)}\int\frac{d^3k}{(2\pi)^3}n_{\chi,k}(t)\sim \frac{1}{a^3(t)}\int\frac{d^3k}{(2\pi)^3}\frac{1}{2}e^{2\pi\mu_k N_c(t)}\\
	&\sim \frac{p_*^{3}}{4\sqrt{2}\pi^2(a(t)/a(t_0))^3\sqrt{\mu_{k_*} N_c(t)}}e^{2\pi\mu_{k_*} N_c(t)}.
	\label{number_density}
\end{align}
We have used the steepest descent method.\footnote{Even with $k$ fixed, the growth factor at each inflaton zero crossing is not the same. Here $\mu_{k_*}$ denotes the value averaged over a number of the crossings.}
Here $t_0$ is the time at the beginning of the broad resonance regime and $N_c(t)$ is the number of the inflaton zero crossings from $t_0$ to $t$.
It is known that $\mu_{k_*}$ is $\mathcal O(0.1)$ in the broad resonance regime. Therefore the typical inverse time scale of the growth of $n_\chi$ may be $\mathcal O(0.1)m_\phi$, which is enough for the modulus to oscillate for our parameter region \eqref{region}.
In our setup, a precise evaluation of Eq.~\eqref{number_density} is difficult because of the anharmonic inflaton oscillation.
Moreover, in the presence of the cosmic expansion, $n_\chi$ grows in a stochastic manner (see Fig.~\ref{fig:oscillation}), which makes the analytic evaluation even more difficult.
Therefore we often have to rely on numerical calculations in order to obtain $n_\chi$ in practice. However, for a special value of $g$ introduced in Sec.~\ref{sec:target}, we can calculate $\langle\chi^2\rangle$ at the beginning of the modulus oscillation without knowing the details of $n_\chi$.

Before going on, let us comment on the modulus production after preheating.
So far we have focused on the modulus abundance induced during preheating. As mentioned at the end of Sec.~\ref{sec:radiation}, the inflaton decay process after preheating in the present model proceeds via some implicitly assumed inflaton interactions. 
Such interactions can in principle induce some amount of modulus abundance if they cause a violent evolution of $\langle\chi^2\rangle$ after preheating. In most cases, however, particle production is most efficient at the early stage of (p)reheating and hence we can safely neglect such effects.

\subsubsection{Target parameter region}
\label{sec:target}

In the above discussion of preheating, we have neglected the back reaction of the radiation on the inflaton dynamics. The back reaction becomes important if the energy density of the radiation $\rho_\chi$ becomes comparable to that of the inflaton $\rho_\phi$. We denote by $g_c$ the value of $g$ for which preheating ends just when $\rho_\chi$ becomes comparable to $\rho_\phi$.
Then the above discussion is reliable until the end of preheating as long as $g\lesssim g_c$.
On the other hand, for $g\gtrsim g_c$, parametric resonance continues in a complicated manner under the efficient back reaction even after $\rho_\chi$ becomes comparable to $\rho_\phi$.
However, as explained later, the resulting modulus abundance for $g\gtrsim g_c$ should not be so different from the one for $g=g_c$. Taking account of it, we investigate the case of $g=g_c$ in the following.
This choice of $g$ makes us possible to compare numerical simulation with the analytical estimation. For other values of $g$, the comparison between numerical simulation and analytic estimation is difficult.

The energy density of $\chi$, Eq.~\eqref{chi_energy}, can be divided into two components:
\begin{align}
	\rho_\chi&=\frac{1}{a^4}\int\frac{d^3k}{(2\pi)^3}\frac{k^2+m_{\chi,\text{eff}}^2}{\omega_k}|\beta_k|^2\\
	&\simeq\frac{1}{a^4}\int\frac{d^3k}{(2\pi)^3}\frac{k^2}{\omega_k}|\beta_k|^2+\frac{g^2\Lambda^2\tanh^2(\varphi/\Lambda)}{a^2}\int\frac{d^3k}{(2\pi)^3}\frac{1}{\omega_k}|\beta_k|^2\\
	&\equiv \rho_{\chi,\text{kin}}+\rho_\text{int},
\end{align}
where we have neglected the terms suppressed by $H^2/g^2\Lambda^2$ in $m_{\chi,\text{eff}}$ from the first to second line.
Here $\rho_{\chi,\text{kin}}$ and $\rho_\text{int}$ denote the kinetic energy of $\chi$ and the interaction energy between $\varphi$ and $\chi$, respectively.
In the broad resonance regime, by using Eq.~\eqref{fluctuation_preheating_first}, the latter is approximated as
\begin{align}
	\rho_\text{int}\sim\frac{g^2\Lambda^2\tanh^2(\varphi/\Lambda)}{2}\langle\chi^2\rangle.
\end{align}
Moreover, in this case, the region $k\ll k_*\sim ag\Lambda$ receives a significant enhancement, leading to $\rho_{\chi,\text{kin}}\ll\rho_\text{int}\simeq\rho_\chi$. With $t_c$ being the time when $\rho_\chi$ becomes comparable to $\rho_\phi$, the fluctuation of $\chi$ at $t=t_c$ is obtained as
\begin{align}
	\langle\chi^2\rangle(t_c)\sim\frac{m_\phi^2}{g_c^2},
	\label{fluctuation_tc}
\end{align}
where we have used $\rho_\chi(t_c)\simeq\rho_\phi(t_c)$. Note that the choice of $g=g_c$ makes it unnecessary to investigate $n_\chi$ in order to obtain $\langle\chi^2\rangle(t_c)$.

Now we can calculate quantities related to the modulus production by applying Eq.~\eqref{fluctuation_tc} to expressions given in Sec.~\ref{sec:modulus}. Here we just give an expression for the energy density of the modulus oscillation, which will be checked by numerical calculation in the next section.
For $t_\text{os}>t_c$, we can take $\langle\chi^2\rangle(t_\text{os})\simeq\langle\chi^2\rangle(t_c)$.
The initial energy density of the modulus oscillation is then obtained from Eqs.~\eqref{sigma_energy} and \eqref{fluctuation_tc} as
\begin{align}
	\rho_\sigma(t_\text{os})
	\simeq\frac{m_\sigma^2}{2}\sigma_\text{min}^2(t_c)
	\simeq\frac{m_\sigma^2}{2}\left(\frac{g_c^2\Lambda^2\tanh^2(\Phi_c/\Lambda)\langle\chi^2\rangle(t_c)}{2Mm_\sigma^2}\right)^2
	\sim \frac{m_\phi^4\Lambda^4\tanh^4(\Phi_c/\Lambda)}{8M^2m_\sigma^2},
	\label{sigma_energy_gc}
\end{align}
where $\Phi_c\equiv\Phi(t=t_c)$ and we have introduced a factor $1/2$ in $\sigma_\text{min}(t_c)$ in order to drop the contribution of the rapidly oscillating term in $\langle\chi^2\rangle$ (see Eq.~\eqref{fluctuation_preheating}).
For $M\sim M_P$, $\rho_\sigma(t_\text{os})\lesssim(H_\text{inf}/m_\sigma)^2\rho_\phi(t_\text{os})\ll \rho_\phi(t_\text{os})$ and the back reaction of $\sigma$ on the background evolution can be neglected.
Note that there is a small correction to the oscillation amplitude if $t_\text{os}<t_c$.

Finally let us consider the case of $g\neq g_c$. For $g\lesssim g_c$, preheating ends before $\rho_\chi$ becomes comparable to $\rho_\phi$. In this case, $\langle\chi^2\rangle$ at the end of preheating is suppressed roughly by a factor of $\rho_\chi/\rho_\phi\,(<1)$ if the system has experienced the broad resonance. 
Without the broad resonance, the suppression becomes more significant.
On the other hand, for $g\gtrsim g_c$, parametric resonance continues still for $t\gtrsim t_c$. In spite of the complicated evolution, the modulus production during the period is expected to be subdominant because $\langle\chi^2\rangle$ as well as $\rho_\chi$ are almost saturated at $t\simeq t_c$. 
Therefore the abundance for $g\gtrsim g_c$ should not be so different from the one for $g=g_c$.

\subsection{Numerical estimation}
\label{sec:numerical}

In this subsection, we numerically investigate the modulus production in the present model.
We calculate $\rho_\sigma(t_\text{os})$ numerically and compare it with the last expression of Eq.~\eqref{sigma_energy_gc}.
In general, the classical lattice simulation may be useful for tracking the time evolution of a system where the back reaction is important. However, since we focus on the case that preheating ends before the back reaction becomes efficient, i.e. $g\lesssim g_c$, it suffices just to solve the differential equations \eqref{eom_inflaton}, \eqref{friedmann}, \eqref{diffeq_bogo} and \eqref{eom_modulus}.

\begin{table}[t]
\centering
\begin{tabular}{|cc||ccccc|c|}
\hline
$\Lambda/M_P$ & $m_\sigma/H_\text{inf}$ & $g_c$ & $\Phi_c/\Lambda$ & $m_\phi/g_c\Lambda$ &$\rho_\chi/\rho_\phi(t_c)$ & $\rho_\text{int}/\rho_\chi(t_c)$ & $\rho_\sigma(t_\text{os})/\eqref{sigma_energy_gc}$\\ \hline
$10^{-2}$ & $20$ & $1.1\times 10^{-3}$ & $0.62$ & $0.65$ & $0.96$ & $0.92$ & $0.43$\\
$10^{-3}$ & $40$ & $4.1\times 10^{-3}$ &$1.3$ & $1.7$ & $0.99$ & $0.88$ & $0.36$\\
$10^{-3}$ & $80$ & $4.1\times 10^{-3}$ &$1.3$ & $1.7$ & $0.99$ & $0.88$ & $0.46$\\
$10^{-3}$ & $160$ & $4.1\times 10^{-3}$ &$1.3$ & $1.7$ & $0.99$ & $0.88$ & $0.18$\\
$10^{-4}$ & $240$ & $1.8\times 10^{-2}$ & $2.0$ & $3.9$ & $0.92$ & $0.84$ & $0.41$\\ \hline
\end{tabular}
\caption{The values of the various quantities in our numerical calculation.
\eqref{sigma_energy_gc} in this table denotes the last expression of Eq.~\eqref{sigma_energy_gc}.
}
\label{tab:value}
\end{table}

\begin{figure}[t]
	\centering
	\includegraphics[width=0.5\linewidth]{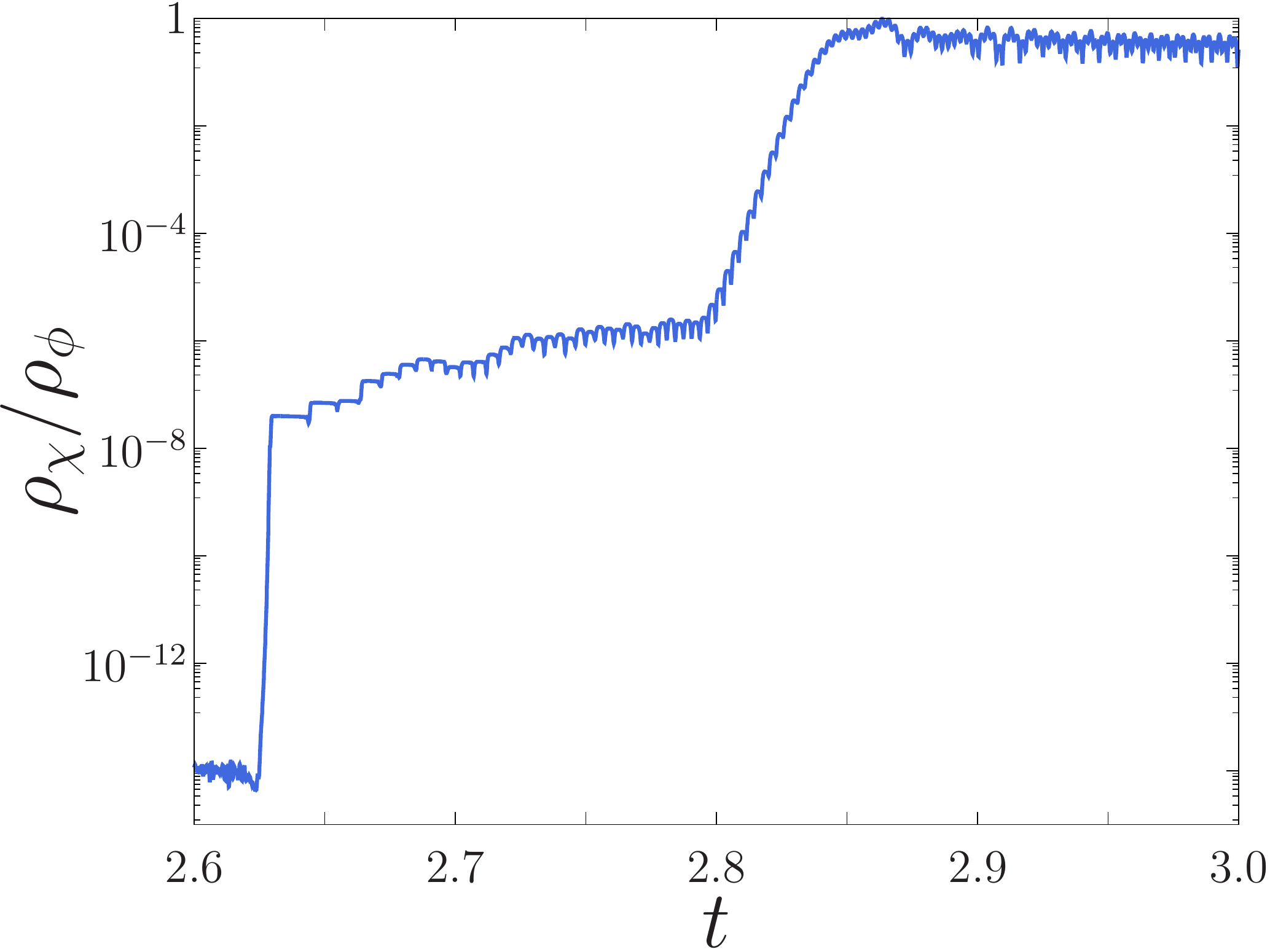}
	\caption{
	\small Time evolution of the ratio between the energy densities of the radiation and the inflaton $\rho_\chi/\rho_\phi(t)$ for $\Lambda/M_P=10^{-3}$ and $g=4.1\times 10^{-3}(=g_c)$. 
	The horizontal axis is $H_\text{inf} t$, and the vertical axis is $\rho_\chi/\rho_\phi$.
Note that the evolution for $H_\text{inf} t\gtrsim 2.85$ is not reliable because our numerical calculation does not include the back reaction.
	}
	\label{fig:ratio}
\end{figure}

As already mentioned, in order to compare the analytic estimation with numerical simulation, it is convenient to take $g=g_c$. 
For given $\Lambda$, we numerically determine $g_c$ via the condition $\rho_\chi(t_c)\simeq\rho_\phi(t_c)$.
In our calculation without the back reaction, $\rho_\chi/\rho_\phi(t)$ becomes maximum after an exponential growth and thereafter it evolves under the cosmic expansion (cf. Fig~\ref{fig:ratio}).
For simplicity, we choose $t_c$ as the moment when $\rho_\chi/\rho_\phi(t)$ becomes maximum.
The values of $g_c$ for various $\Lambda$ are shown in Table~\ref{tab:value}.\footnote{
	For these values of $g_c$, the radiative correction to the inflaton potential is negligible.
}
With $g=g_c$, we calculated $\rho_\sigma(t_\text{os})$ for various $\Lambda$ and $m_\sigma$ without neglecting terms suppressed by a factor of $H^2/g^2\Lambda^2$.

In Fig.~\ref{fig:ratio}, we show the time evolution of the ratio between the energy densities of the radiation and the inflaton $\rho_\chi/\rho_\phi(t)$ for $\Lambda/M_P=10^{-3}$ and $g=4.1\times 10^{-3}(=g_c)$.
The initial condition is the same as the one for Fig.~\ref{fig:oscillation}.
It is seen that $\rho_\chi/\rho_\phi$ receives enhancements in a stochastic manner like $n_\chi$ in the right figure of Fig.~\ref{fig:oscillation}. Because of the definition of $g_c$, the parametric resonance ceases and $\rho_\chi/\rho_\phi$ is almost constant after $\rho_\chi/\rho_\phi$ becomes around unity. 

\begin{figure}
	\begin{minipage}{0.5\hsize}
		\centering
		\includegraphics[width=\linewidth]{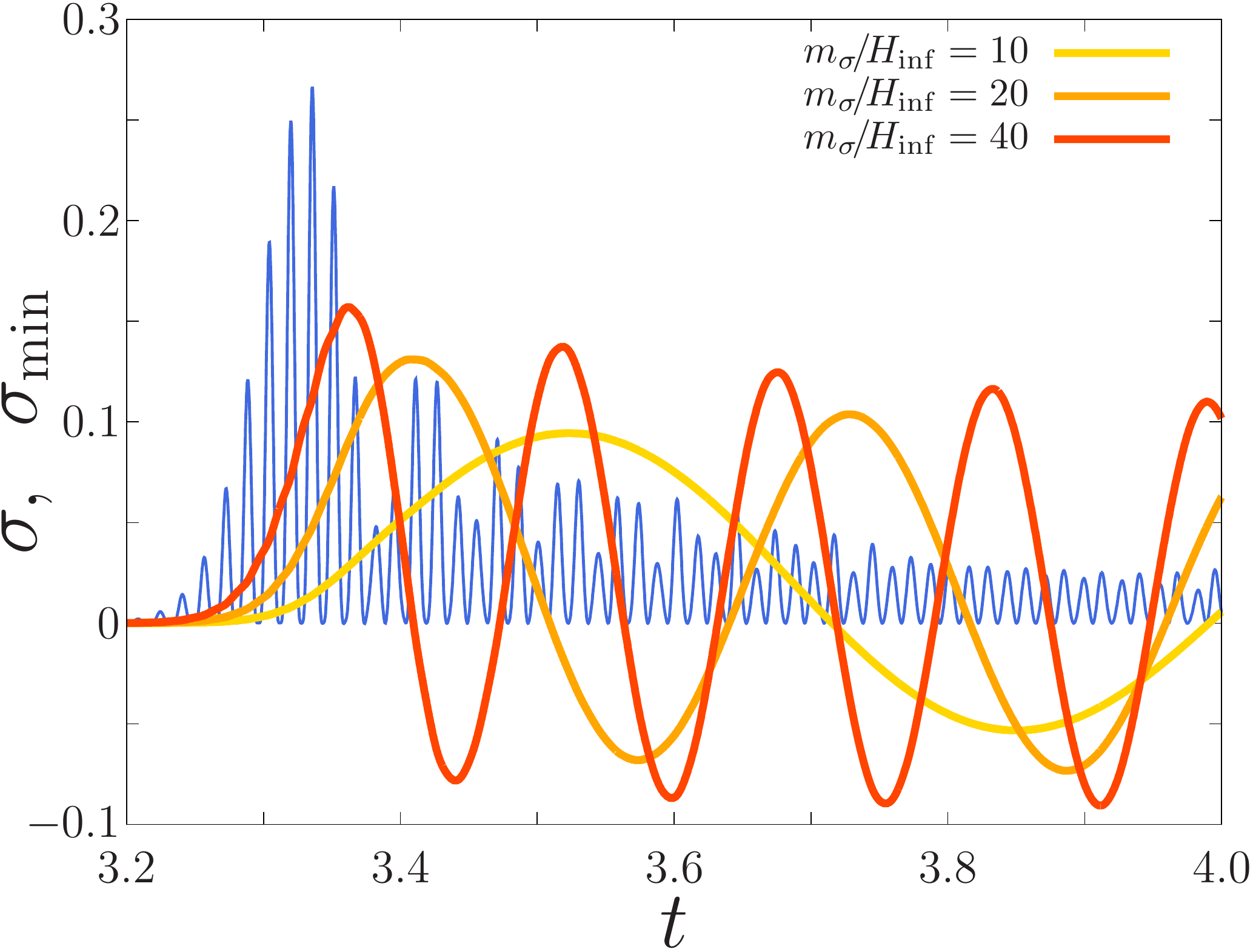}
	\end{minipage}
	\begin{minipage}{0.5\hsize}
		\centering
		\includegraphics[width=\linewidth]{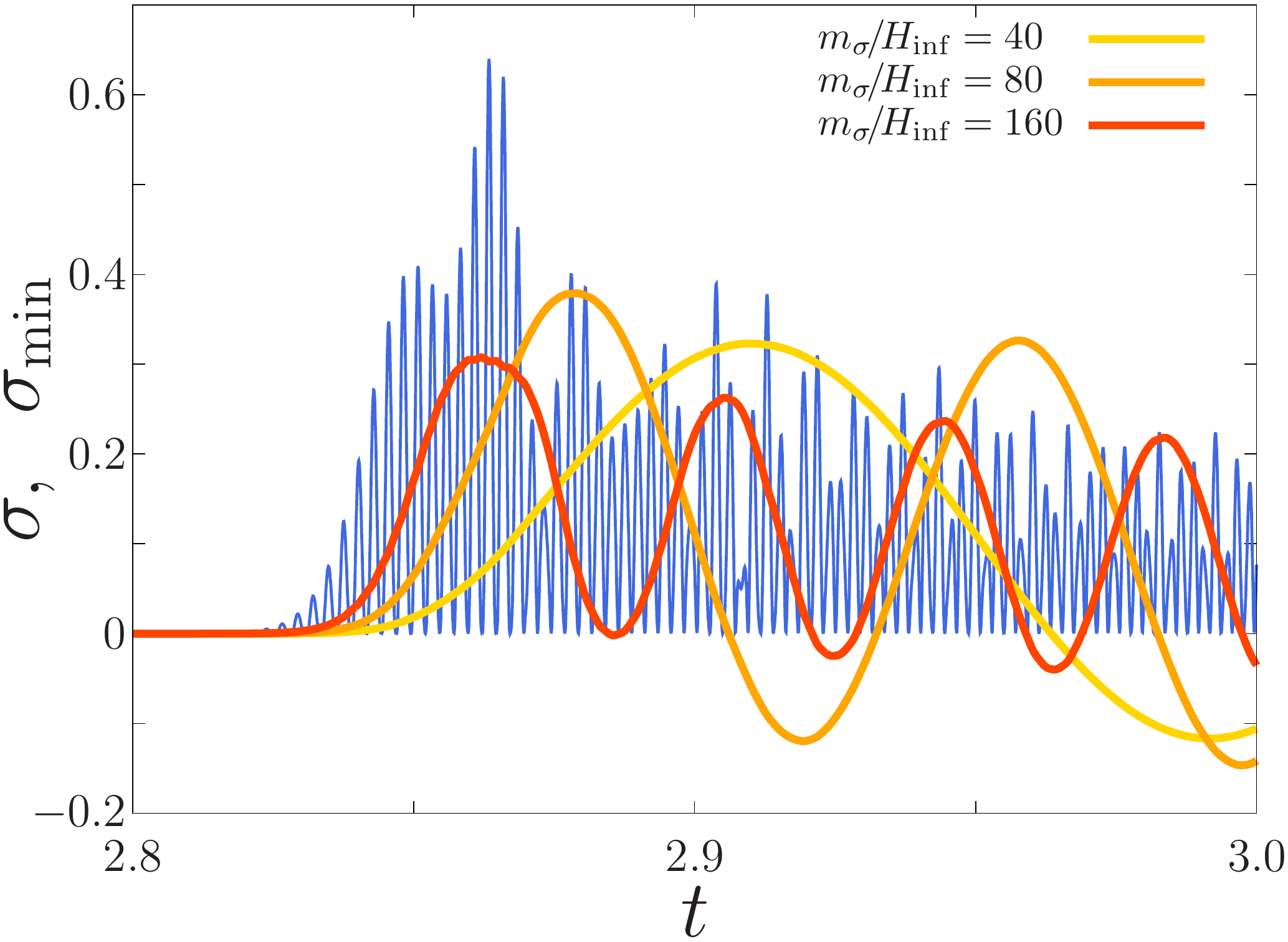}
	\end{minipage}
	\begin{minipage}{0.5\hsize}
		\centering
		\includegraphics[width=\linewidth]{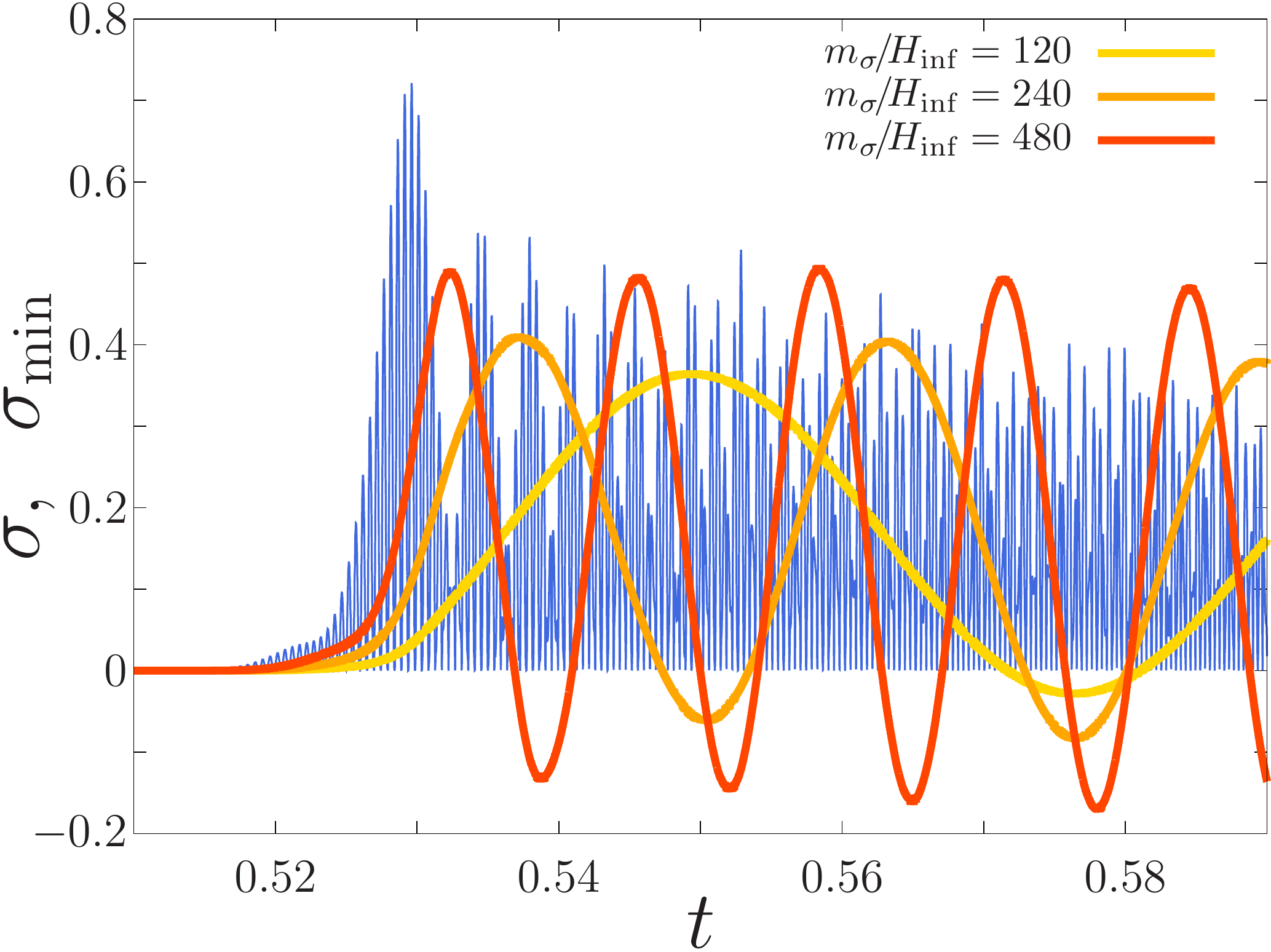}
	\end{minipage}
	\begin{minipage}{0.5\hsize}
		\centering
		\includegraphics[width=\linewidth]{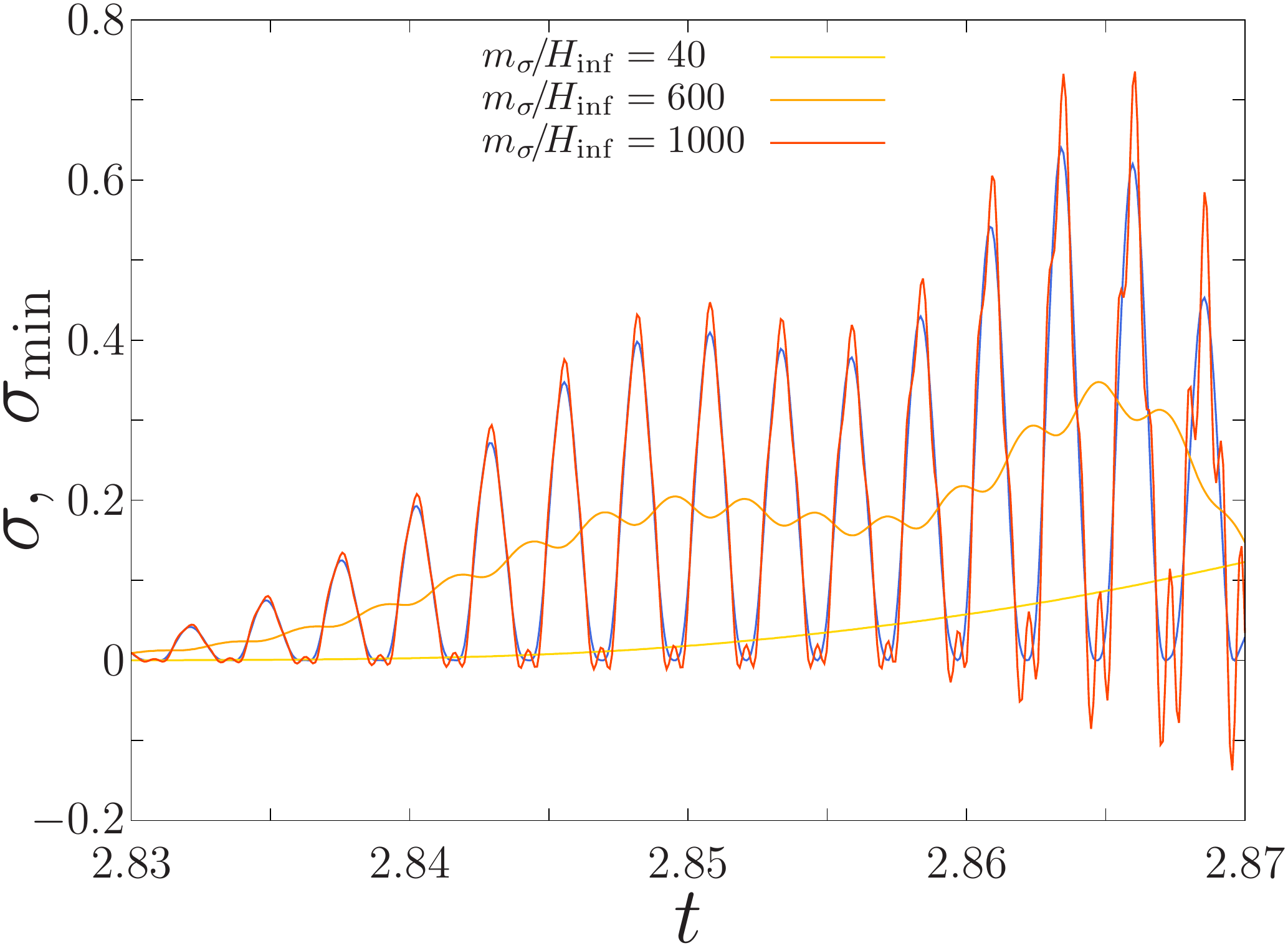}
	\end{minipage}
	\caption{
	\small Time evolutions of the modulus position $\sigma(t)$ and the modulus potential minimum $\sigma_\text{min}(t)$ (\textcolor{blue}{blue}) for various values of $\Lambda$ and $m_\sigma$ with $g=g_c$. The horizontal axis is time $t$ in unit of $H_{\rm inf}^{-1}$. 
The vertical axis is $\sigma$ or $\sigma_\text{min}$ in unit of $m_\phi^2\Lambda^2/m_\sigma^2M$.
The upper left and lower left figures are for $\Lambda/M_P=10^{-2}~\text{and}~10^{-4}$, respectively. The upper right and lower right figures are for $\Lambda/M_P=10^{-3}$.
The lower right figure is for very heavy moduli.
	}
	\label{fig:potential}
\end{figure}

The time evolutions of the modulus position $\sigma(t)$ and the modulus potential minimum are shown in Fig.~\ref{fig:potential}.
The initial position of the inflaton is chosen as $\varphi(t)=2.0\varphi_\text{end}~(1.8\varphi_\text{end})$ for $\Lambda/M_P=10^{-2},10^{-3}~(10^{-4})$. The corresponding initial velocity of the inflaton is derived from Eqs.~\eqref{eom_inflaton} and \eqref{friedmann} with slow-roll approximation $\varphi''(0)=0$. On the other hand, we take $(\alpha_k,\beta_k)=(1,0)$ for all $k$ and $(\sigma,\sigma')=(0,0)$ at a moment slightly before the first inflaton zero crossing, though different choices of the moment do not change the results much.
The modulus oscillation is induced for $H_\text{inf}\ll m_\sigma\ll m_\phi$ because the modulus cannot keep up with the quick change of the potential minimum during preheating.

In Table.~\ref{tab:value}, we show the ratio of the numerically obtained modulus energy density to the analytical one, 
$\rho_\sigma(t_\text{os})/\eqref{sigma_energy_gc}$, together with various quantities in our numerical calculation.
Here $\rho_\sigma(t_\text{os})$ is evaluated as $m_\sigma^2[(\sigma_1+\sigma_2)/2]^2/2$ where $\sigma_1$ ($\sigma_2$) is the value of the modulus position at the first positive (negative) peak of the induced modulus oscillation.
The analytical estimation agrees with the numerical results within an order of magnitude.
$\rho_\sigma(t_\text{os})/\eqref{sigma_energy_gc}$ slightly depends on $m_\sigma$, which may reflect the non-exact relation $\hat{\sigma}(t_\text{os})\simeq\sigma_\text{min}(t_c)$. 
Both $\Phi_c/\Lambda$ and $m_\phi/g_c\Lambda$ are comparable to or larger than unity, which are utilized in the discussion of Sec.~\ref{sec:preheating}.
$\rho_\text{int}/\rho_\chi(t_c)\sim1$ justifies the approximation used in Sec.~\ref{sec:target}.

These numerical results confirm our analytical estimation in Sec.~\ref{sec:analytical}.
The resulting modulus abundance shows that this new modulus production mechanism is so efficient that it can be cosmologically harmful in general.
It is notable that comparable amount of modulus abundance may be induced also for a rather wide range of the coupling constant $g\gtrsim g_c$.

Before closing this subsection, we make brief comments on the case of very heavy moduli and the limitation of our numerical calculation.
First, so far we have focused on the region \eqref{region}: $H_\text{inf}\ll m_\sigma\ll m_\phi$.
If $m_\sigma$ becomes close to $m_\phi$, the condition \eqref{sigma_condition} may not be satisfied any more and the oscillation may be significantly suppressed.
Examples of such a behavior can be seen in the lower right figure of Fig.~\ref{fig:potential}.
The figure shows that, if $m_\sigma$ is very large and comparable to $m_\phi$, the modulus keeps tracking the potential minimum and the oscillation is significantly suppressed. The line for $m_\sigma/H_\text{inf}=600$ tracks the potential minimum without feeling its rapidly oscillating part (see Eq.~\eqref{fluctuation_preheating}) while the one for $m_\sigma/H_\text{inf}=1000$ tracks the potential minimum including it.
Second, we have shown results only for $10^{-4}\leq\Lambda/M_P\leq10^{-2}$. We found that for $\Lambda/M_P\ll 10^{-4}$ the condition $\rho_\chi(t_c)\simeq\rho_\phi(t_c)$ is satisfied without the broad resonance in our setup. Though the modulus oscillations are still induced even in this case, their abundances are suppressed compared to the analytical expressions given in Sec.~\ref{sec:analytical} and it is difficult to provide their rigorous estimations.

\section{Summary and discussion}
\label{sec:summary}

In this work we have studied in detail the coherent oscillation of the moduli whose mass is much larger than the Hubble scale during inflation: $m_\sigma \gg H_{\rm inf}$. We pointed out that heavy moduli linearly coupled to some operator like Eq.~\eqref{linearterm} can have sizable abundance in the form of the coherent oscillation, since it induces a rapid shift of the effective moduli potential during the reheating regime. The resulting abundance is in general so large that cosmological moduli problem may not be solved by simply making the moduli heavy. 
As an illustrative example, we considered the $\alpha$-attractor inflation coupled to light sector, in which particle production occurs efficiently after inflation and as a result the modulus potential is affected in a time-dependent and non-trivial way.  By numerically solving the equations of motion, it is shown that the moduli oscillation is induced at the very beginning of the reheating stage. 
In this example we have only considered the regime in which the back reaction of the light sector to the inflaton is neglected and also the light sector is only weakly interacting so that they are not thermalized.
Although we need further detailed simulations to incorporate these effects, in any case it is likely that the operators $\mathcal O$ to which the modulus couples are highly time-dependent and the modulus oscillation is induced accordingly, unless the modulus mass is much larger than any other inverse time scale of the dynamics.
Therefore, in order to avoid the efficient modulus production, the modulus mass needs to be at least much larger than the inflaton: $m_\sigma \gg m_\phi$.

We emphasize that the production mechanism derived in this paper is applied to broad class of scalar fields.
In particular, the modulus $\sigma$ can be identified as some symmetry breaking field such as saxions in supersymmetric axion models or flavons in models with flavor symmetry, and so on, since they are necessarily linearly coupled to some operators like Eq.~(\ref{linearterm}) around their potential minimum once the cutoff scale $M$ is identified with (some coupling constant times) the symmetry breaking scale. One caution is that the correction to the potential can be so large that the modulus $\sigma$ may be shifted more than $M$. In such a case, we need to go beyond the expansion around the vacuum and the dynamics becomes much more complicated. 

Finally, we note that there are several other production mechanisms of the moduli.
Unless the moduli are conformally coupled to the Ricci scalar ($\xi \neq 1/6$), moduli feel the time-dependent background and the so-called gravitational particle production happens~\cite{Parker:1969au,Ford:1986sy,Chung:1998zb,Chung:2001cb}. This works for the mass range we are interested in: $H_{\rm inf} < m_\sigma < m_\phi$~\cite{Ema:2015dka,Ema:2016hlw,Ema:2018ucl}.
Thermal scatterings can also produce moduli if the temperature of the universe ever exceeds the mass of the moduli. These contribute to the momentum distribution of the moduli at the scale around the inflaton mass or the temperature, and they are distinguished from the coherent oscillation studied in this paper.
When discussing the cosmological effects of moduli, one should take into account both these high momentum contributions and coherent oscillation even if the moduli are heavier than the Hubble scale during inflation.

\section*{Acknowledgments}

This work was supported in part by the Grant-in-Aid for Scientific Research A (No.\ 16H02189 [KH]), Scientific Research C (No.\ 18K03609 [KN]), and Innovative Areas (No.\ 26104001 [KH], No.\ 26104009 [KH and KN], No.\ 15H05888 [KN], No.\ 17H06359 [KN]).

\bibliographystyle{JHEP}
\bibliography{ref}

\end{document}